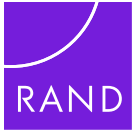

EDWARD PARKER, NICHOLAS A. O'DONOUGHUE, ALVIN MOON, NICOLAS M. ROBLES

# Assessing the Practical Feasibility of the Clader-Jacobs-Sprouse Quantum Algorithm for Calculating Radar Cross Sections

Research Report

For more information on this publication, visit **www.rand.org/t/RRA4086-1**.

**About RAND**

RAND is a research organization that develops solutions to public policy challenges to help make communities throughout the world safer and more secure, healthier and more prosperous. RAND is nonprofit, nonpartisan, and committed to the public interest. To learn more about RAND, visit www.rand.org.

**Research Integrity**

Our mission to help improve policy and decisionmaking through research and analysis is enabled through our core values of quality and objectivity and our unwavering commitment to the highest level of integrity and ethical behavior. To help ensure our research and analysis are rigorous, objective, and nonpartisan, we subject our research publications to a robust and exacting quality-assurance process; avoid both the appearance and reality of financial and other conflicts of interest through staff training, project screening, and a policy of mandatory disclosure; and pursue transparency in our research engagements through our commitment to the open publication of our research findings and recommendations, disclosure of the source of funding of published research, and policies to ensure intellectual independence. For more information, visit www.rand.org/about/research-integrity.

RAND's publications do not necessarily reflect the opinions of its research clients and sponsors.

# About This Report

Shor's algorithm, which could allow quantum computers to solve cryptographic problems that are intractable for classical computers, has led to decades of intense research into realizing quantum computers physically and finding other quantum algorithms that provide *exponential speedups* relative to their classical counterparts.

In 2008, Aram Harrow, Avinatan Hassidim, and Seth Lloyd discovered an algorithm with the potential for such speedup when solving certain linear systems of equations, and in 2013, David Clader, Bryan Jacobs, and Chad Sprouse developed an extension of that algorithm, the Clader-Jacobs-Sprouse (CJS) algorithm, that solves electromagnetic scattering problems and demonstrates an end-to-end exponential speedup over classical algorithms for the same problem. If quantum computers of sufficient size are realized, then this CJS algorithm could theoretically solve many radio frequency problems, such as the modeling of a target's radar cross section, much more rapidly than is possible as of 2026. This would be important for modeling and predicting radar behavior against emerging targets, such as unmanned aerial systems, small satellites, space debris, and adversary aircraft (such as low-observable fighter and bomber aircraft in development).

In this report, we compare the end-to-end computational complexity and resource costs of the CJS algorithm (including the read-in and read-out steps that are not always analyzed in the quantum algorithms research literature) with the closest classical approach to the same problem (the frequency-domain finite-element-method with a conjugate gradient solver). We assess the likelihood of a practically useful quantum advantage from the CJS algorithm.

This report is intended for consumption by technical professionals with familiarity with either quantum algorithms or radar and for decisionmakers who are responsible for funding research into such systems. It is written to minimize or explain all technical jargon.

## RAND National Security Research Division

This research was conducted within the Acquisition and Technology Policy Program of the RAND National Security Research Division (NSRD), which operates the RAND National Defense Research Institute (NDRI), a federally funded research and development center (FFRDC) sponsored by the Office of the Secretary of War, the Joint Staff, the Unified Combatant Commands, the Navy, the Marine Corps, the defense agencies, and the defense intelligence enterprise.

For more information on the RAND Acquisition and Technology Policy Program, see www.rand.org/nsrd/atp or contact the director (contact information is provided on the webpage).

This publication documents work completed in December 2025, which underwent security review with the sponsor and the Defense Office of Prepublication and Security Review before public release.

## Funding

This research was made possible by NDRI exploratory research funding that was provided through the FFRDC contract and approved by NDRI's primary sponsor.

## Acknowledgments

This report was funded by internal research support provided by RAND's National Security Research Division. We would like to thank the individuals who provided this support, including director Barry Pavel and associate director Anu Narayanan; Caitlin Lee and Megan McKernan, the directors of the Acquisition and Technology Policy program within the division; and the support personnel who were instrumental in executing this work—specifically, Taria Francois and Megan Hasak.

We would like to thank the members of the quality assurance panel, Carter Price and Anupam Prakash, and the quality assurance manager, Lance Menthe, for their thoughtful and constructive review of this report. Nevertheless, any errors that remain are the sole responsibility of the authors.

# Summary

## Issue

Many policy-relevant problems could benefit from breakthrough advances in computing capability. One of those challenging computational problems concerns calculating *radar cross sections* (RCSs) off a given target, which has such applications as designing the next generation of low-observable aircraft and predicting the performance of various sensors against a wide variety of threats. Understanding which approaches offer the most technical promise over the short and the long terms can help policymakers allocate scarce research and development resources across those approaches. Moreover, understanding the promise and likely timelines for these potential breakthroughs can help force designers to prepare for various futures in which U.S. forces (or their adversaries) gain new breakthrough military capabilities.

The new class of computers known as *quantum computers* offers promise for a variety of transformative applications with major implications for both economic and national security. Quantum computers execute special *quantum algorithms* that are very different from standard computer algorithms. Since Peter Shor discovered his famous quantum algorithm for fast factoring in 1994, there has been an enormous research effort to discover new quantum algorithms.

The most-promising quantum algorithms offer an *exponential speedup* over the best known classical algorithms for the same computational problems. These algorithmic speedups are so dramatic that they could overcome the huge hardware overhead required to operate a large-scale quantum computer. Unfortunately, relatively few known quantum algorithms offer a clear exponential speedup over the fastest corresponding classical algorithms. Some of those algorithms, such as Shor's algorithm for factoring and Lloyd's algorithm for simulating quantum systems, have been extensively studied.

But another such algorithm has been much less well studied: the Clader-Jacobs-Sprouse (CJS) algorithm for calculating RCSs for a given target. Even though researchers are still very far from being able to build a quantum computer capable of running the CJS algorithm, policymakers need to understand the promise and pitfalls of such quantum algorithms to properly allocate research and acquisition resources between improvements to existing approaches to RCS modeling and fundamentally new quantum-computing–based approaches.

## Approach

In this report, we analyze the computational resources that a hypothetical quantum computer is predicted to require to run the CJS algorithm more efficiently than a classical computer could perform a similar computation. The main computational resource that we consider is the runtime, but we also touch on the memory requirements (i.e., the number of qubits required to run the algorithm on a useful problem size). We first perform an *asymptotic analysis*, which focuses on the (somewhat coarse) functional scaling of the runtime with problem size, and then we analyze and extend a quantitatively precise resource estimate from the previous literature. The main theme of our analysis is to try to conceptually understand the key problem parameters and regimes in which either the classical or the quantum approach is likely to dominate the other.

## Findings

Our main finding is that because of its exponential speedup over comparable classical algorithms, the CJS algorithm is a somewhat promising candidate practical application for eventual large-scale quantum computers—perhaps one of the most promising candidate applications after the simulation of the physics of quantum systems.

But we also find several important caveats that reduce the algorithm's promise. For example, we find that even though the algorithm can, in principle, efficiently solve very large problem sizes (corresponding to a discretization of the spatial region being modeled into a very fine computational mesh), its runtime is highly sensitive to various other problem parameters, such as the mesh topology, and all three-dimensional mesh topologies impose very high resource costs from these other parameters. We discuss which parameter choices are best tailored to make the CJS algorithm run as efficiently as possible.

Another caveat is that there is no rigorous proof or even strong evidence from computational complexity theory that the CJS algorithm is faster than the fastest *possible* classical algorithm; it is possible that a researcher could discover a new classical algorithm for RCS calculations that matches the CJS algorithm's performance.

Finally, concrete computational resource estimates from the literature suggest that the resources required to run the CJS algorithm in practice could be *far* higher than an asymptotic analysis might imply. We combine these previous estimates with some plausible hardware estimates for future quantum computers to reach the following *very* rough back-of-the-envelope estimate: Running the specific implementation of the CJS algorithm modeled in the literature to model even a two-dimensional toy RCS scattering problem would require 100 million times the age of the universe. This calculation would obviously not be feasible in practice.

All in all, we conclude that the CJS algorithm is significantly less promising than the simulation of quantum systems as a near-term application of quantum computers. Therefore, research funders should bear in mind that despite quantum computers' apparent promise for RCS modeling, they are unlikely to deliver breakthrough capabilities in that area in the near term.

But we do not believe that research funders or force designers should write off the CJS algorithm completely. We identify several promising concrete pathways for future algorithmic research that could greatly decrease these resource costs, possibly all the way down to the regime of practical feasibility. For example, we identify a fairly new computational subroutine (the Childs-Kothari-Somma algorithm) that could potentially alleviate a key bottleneck from the existing CJS algorithm and greatly speed up its runtime.

We find that the critical bottleneck for the CJS algorithm is a quantum subroutine known as *Hamiltonian simulation*, which is more often studied in the context of simulating the physics of quantum systems than in the context of calculating RCSs. Therefore, we reach the surprising conclusion that a breakthrough in the modeling of RCSs could emerge from a research effort with a completely different goal, such as medical drug discovery. We recommend that quantum computing funders and the RCS modeling community monitor scientific progress in the field of Hamiltonian simulation (including by biomedical researchers) and consider how this progress might also accelerate the timelines for utility of the CJS algorithm.

From this relatively short study, we have identified several concrete open questions that will require further research before we can confidently assess the practical feasibility of the CJS algorithm. We recommend that funders of quantum research who want to advance the state of the art for RCS modeling prioritize these research questions and pathways. Although the CJS algorithm seems to be far from the most promising near-term quantum algorithm, it is still too soon to rule it out entirely.

# Contents

# Figures and Table

## Figures

## Table

# Chapter 1

# Introduction and Motivation

## Background

The radar range equation computes the received power level from a transmitted signal as a function of many parameters, most notably an object's radar cross section (RCS). The RCS has both military and civilian applications and can be used for many tasks, such as setting system requirements, predicting system performance, planning operations, associating detections to form a track, rejecting false or spurious targets, and classifying detected objects.

These tasks can apply to military problems, such as the detection of ballistic missiles in an early warning system, imaging target areas with Synthetic Aperture Radar (SAR) from a reconnaissance platform or guiding a missile to its intended target. Closely related civilian missions include orbital-debris tracking for space domain awareness and collision avoidance and the use of SAR imagery for mapping (e.g., land-use data, scientific observation). These tasks can equally apply to the operation of radar modules supporting adaptive cruise control, automated collision avoidance systems in drones or autonomous vehicles, or mapping and navigation sensors in robotic devices.

To resolve any of these tasks, it is critical to estimate or understand the object's RCS—the strength of the reflection from it. This value will vary with many parameters, including the frequency of the incoming waves, the orientation of the object with respect to those waves, and the polarization of the waves. There are also many environmental factors that can affect RCS (such as the presence of moisture on the surface of an object), but we will ignore those for the sake of simplicity in this report.

The prediction of a target's RCS begins with a high-fidelity model of the target's structure, including surface material compositions, paints or appliqués, and (if sufficient accuracy is desired) manufacturing tolerances and fastener types. These are typically fed into an electromagnetic (EM) solver, which may require hours or days for a single run. The computation of a full RCS file will require many repeated runs at different frequencies and aspect angles and potentially with different configurations (e.g., with payload bay doors open or closed). The computational resources required can be quite immense.

To add to the challenge, there has been tremendous growth in the number of targets for which an RCS prediction would be useful, including drones (which may be upgraded rapidly during conflict),[1] new adversary aircraft, small satellites, and space debris.[2] If it is feasible, the exponential speedup promised by quantum algorithms for the prediction of RCS would have tremendous value.

---

[1] Todd Prince, "Pentagon Faces 'Wake-Up Call' to Meet Drone Innovation Highlighted in Ukraine War," Radio Free Europe, August 7, 2025.

[2] Justin K. A. Henry, Ram M. Narayanan, and Puneet Singla, "Radar Cross-Section Modeling of Space Debris," in Erik Blasch, Frederica Darema, and Alex Aved, eds., *Dynamic Data Driven Applications Systems: 4th International Conference, DDDAS 2022, Cambridge, MA, USA, October 6–10, 2022, Proceedings*, Springer Cham, 2024.

# Quantum Computing Algorithms That Deliver an Exponential Speedup over Classical Algorithms

In 1982, the physicist Richard Feynman proposed a fundamentally new type of computer: a *quantum computer*, which would potentially take advantage of the exotic properties of quantum physics to perform certain calculations far faster than standard digital computers could.[3] At the most fundamental level, standard (or *classical*) computers can be described using an idealized theoretical model, known as a Turing machine. But a quantum computer would operate on completely different principles and could *not* be modeled as a Turing machine.

To illustrate just how different quantum computers are from conventional computers, consider that quantum computers do not operate using bits, the familiar 0s and 1s that underlie all of a classical computer's lowest-level data representation and processing. Instead, a quantum computer would store and operate on data represented as *qubits* (or "quantum bits"), which can exist in a *quantum superposition* of 0 and 1 at the same time. Rather than the standard deterministic operations on bits (such as the AND, OR, and NOT operations), quantum computers manipulate *probability amplitudes* of various bitstrings (which are related to but not identical to conventional probabilities) according to certain mathematical rules. These rules are believed to allow quantum computers to perform certain calculations vastly faster than is possible by the classical rules of Boolean arithmetic.

A quantum computer requires two ingredients: quantum hardware and software (i.e., algorithms). Neither is easy to develop.

Since Feynman's 1982 proposal, quantum hardware has advanced by an enormous amount, and quantum computers are no longer theoretical. In 2019, the first quantum computer crossed a milestone known as *quantum supremacy*: It performed a proof-of-concept computation in five minutes that, at the time, was estimated to require 10,000 years on the world's fastest supercomputer.[4] There are many different hardware approaches to developing quantum computers.[5] But for the purpose of this report, the key point is that quantum computers perform sequential logical operations on data that are stored in a large number of qubits. (The total number of qubits is roughly analogous to the amount of memory used by a classical computer.)

The second necessary ingredient is a *quantum algorithm* that respects the rules of quantum computing to perform a useful computation. There are dozens of known quantum algorithms, but they fall into a small number of major categories.[6]

---

[3] Richard P. Feynman, "Simulating Physics with Computers," *International Journal of Theoretical Physics*, Vol. 21, Nos. 6–7, June 1982.

[4] Frank Arute, Kunal Arya, Ryan Babbush, Dave Bacon, Joseph C. Bardin, Rami Barends, Rupak Biswas, Sergio Boixo, Fernando G. S. L. Brandao, David A. Buell, et al., "Quantum Supremacy Using a Programmable Superconducting Processor," *Nature*, Vol. 574, No. 7779, October 23, 2019. The question of whether classical supercomputers could actually match that computer's performance is quite complicated and played out over several years after the 2019 demonstration. But by 2024, there was very strong evidence that quantum computers could perform certain computations that were far beyond the capability of classical supercomputers (A. Morvan, B. Villalonga, X. Mi, S. Mandrà, A. Bengtsson, P. V. Klimov, Z. Chen, S. Hong, C. Erickson, I. K. Drozdov, et al., "Phase Transitions in Random Circuit Sampling," *Nature*, Vol. 634, No. 8033, October 9, 2024).

[5] T. D. Ladd, F. Jelezko, R. Laflamme, Y. Nakamura, C. Monroe, and J. L. O'Brien, "Quantum Computers," *Nature*, Vol. 464, No. 7285, March 4, 2010.

[6] Stephen P. Jordan, "Quantum Algorithm Zoo," last updated March 31, 2025.

Most known quantum algorithms only deliver a *polynomial speedup* (usually quadratic) over the best known classical algorithms for the same problem. That is, if the runtimes of the best known classical algorithm and the best known quantum algorithm scale in the size of the problem input $n$ as $c_n$ and $q_n$, respectively, then $c_n$ only grows as a polynomial in $q_n$. Given the huge hardware overheads required to operate a quantum computer, many experts are skeptical that a polynomial speedup for a given problem will be sufficient to make a quantum computer more useful in practice than a classical computer for that problem.[7]

A much more promising path for quantum computers to deliver a useful advantage over classical computers is through algorithms that deliver an *exponential* speedup over the best known classical algorithm; that is, $c_n$ grows exponentially fast in $q_n$. In this case, the quantum speedup is so dramatic that it may overcome the huge hardware overhead required by quantum computers.

There are many quantum algorithms that are believed to offer an exponential speedup over classical algorithms, but unfortunately, many of them solve esoteric math problems with little practical utility. There are only a few known categories of quantum algorithm for *useful* problems that give an exponential speedup over the best known classical algorithm for the same problem. The first is *Shor's algorithm* for rapidly factoring large numbers, which could have important implications for modern cryptography.[8] The second category is a collection of algorithms for modeling the time evolution of quantum systems themselves, which are typically variations on an algorithm discovered by Seth Lloyd.[9]

The third category of algorithms are those used to solve large systems of linear equations. That is, given a large matrix $A$ and vector $\boldsymbol{b}$ satisfying certain conditions, the task is to solve the matrix equation $A\boldsymbol{x} = \boldsymbol{b}$ for the unknown vector $\boldsymbol{x}$. These algorithms are known as (variations of) the *quantum linear system algorithm* (QLSA), and they are the focus of this report. The basic algorithm was discovered by Aram Harrow, Avinatan Hassidim, and Seth Lloyd (HHL) in 2008.[10]

As we will discuss in Chapter 2, these algorithms are bit more conceptually subtle than Shor's or Lloyd's algorithms, and their practical utility is not entirely clear. Most versions of the QLSA face severe bottlenecks in both uploading the problem data into the quantum processor and reading them back out again. In most cases, these input/output bottlenecks may eliminate the quantum speedup, so the true utility of the QLSA is not entirely clear.[11]

---

[7] Ryan Babbush, Jarrod R. McClean, Michael Newman, Craig Gidney, Sergio Boixo, and Hartmut Neven, "Focus Beyond Quadratic Speedups for Error-Corrected Quantum Advantage," *PRX Quantum*, Vol. 2, No. 1, March 29, 2021; Torsten Hoefler, Thomas Häner, and Matthias Troyer, "Disentangling Hype from Practicality: On Realistically Achieving Quantum Advantage," *Communications of the ACM*, Vol. 66, No. 5, May 2023.

[8] Peter W. Shor, "Polynomial-Time Algorithms for Prime Factorization and Discrete Logarithms on a Quantum Computer," *SIAM Journal on Computing*, Vol. 26, No. 5, 1997.

[9] Seth Lloyd, "Universal Quantum Simulators," *Science*, Vol. 273, No. 5278, August 23, 1996.

[10] Aram W. Harrow, Avinatan Hassidim, and Seth Lloyd, "Quantum Algorithm for Linear Systems of Equations," *Physical Review Letters*, Vol. 103, No. 15, October 9, 2009.

[11] Scott Aaronson, "Read the Fine Print," *Nature Physics*, Vol. 11, No. 4, April 2, 2015.

## Quantum Computing Algorithms for Solving Differential Equations

One important class of application for QLSA-type linear solvers is for numerically solving partial differential equations (PDEs). In this section, we briefly point the reader toward some of the key literature on this topic without going into detail. In the rest of this report, we will focus in more detail on one particular quantum algorithm that uses the QLSA as a subroutine to solve Maxwell's PDEs.

Early work analyzed when finite-element discretization methods can benefit from quantum linear-system solvers and highlighted limitations because of conditioning and readout overhead.[12] Building on that, algorithms for linear differential equations achieved exponentially improved scaling.[13] These were extended to other PDE solver algorithms that used spectral and adaptive finite-difference schemes, which are much more accurate than the previous algorithms.[14] Quantum finite-element implementations now incorporate multilevel preconditioning,[15] and they can solve multiscale PDEs.[16]

Other work has explored variational approaches for solving PDEs.[17] These approaches are generally less accurate than linear solvers, but they have much lower hardware requirements and could potentially be run on near-term quantum computers. They provide heuristic solvers for elliptic PDEs, such as the Poisson equation.

Yet other recent methods use *Hamiltonian simulation* or *Schrödingerization* methods to recast linear PDEs as time-evolution problems that are amenable to simulation.[18]

Lastly, the analog of the finite-element method (FEM) for stochastic partial differential equations (SDEs) is known as the *Euler-Maruyama method*. Certain properties of SDEs can be mapped to PDEs

---

[12] Ashley Montanaro and Sam Pallister, "Quantum Algorithms and the Finite Element Method," *Physical Review A*, Vol. 93, No. 3, March 2016.

[13] B. D. Clader, B. C. Jacobs, and C. R. Sprouse, "Preconditioned Quantum Linear System Algorithm," *Physical Review Letters*, Vol. 110, No. 25, June 2013; Dominic W. Berry, Andrew M. Childs, Aaron Ostrander, and Guoming Wang, "Quantum Algorithm for Linear Differential Equations with Exponentially Improved Dependence or Precision," *Communications in Mathematical Physics*, Vol. 356, No. 3, December 2017.

[14] Andrew M. Childs, Jin-Peng Liu, and Aaron Ostrander, "High-Precision Quantum Algorithms for Partial Differential Equations," *Quantum*, Vol. 5, November 10, 2021. By *much more accurate*, we mean that PDEs' runtime grows only polylogarithmically in the inverse error.

[15] Matthias Deiml and Daniel Peterseim, "Quantum Realization of the Finite Element Method," *Mathematics of Computation*, August 14, 2025. Technical note: this work implements the type of conditioning discussed in James H. Bramble, Joseph E. Pasciak, and Jinchiao Xu, "Parallel Multilevel Conditioners," *Mathematics of Computation*, Vol. 55, No. 191, 1990: multilevel additive schemes that combine information from coarse and fine finite-element grids to reduce the condition number of the stiffness matrix. If $A_l$ is the finite-element stiffness matrix on level $l$, with prolongation operators $P_l$ and local smoothers $D_l \approx A_l$, the preconditioner

$$B^{-1} = \sum_{l=0}^{L} P_l D_l^{-1} P_l^T$$

yields $\kappa(B^{-1} A_L) = O(1)$, making convergence essentially mesh-independent. Such BPX-type methods are important for quantum FEM solvers because the runtime of quantum linear-systems algorithms scales with $\kappa(A_L)$.

[16] Junpeng Hu, Shi Jin, and Lei Zhang, "Quantum Algorithms for Multiscale Partial Differential Equations," arXiv, arXiv:2304.06902, April 14, 2023.

[17] Yilian Liu, *A Variational Quantum Algorithm for Solving Partial Differential Equations*, thesis, Cornell University, May 2023.

[18] Liu, 2023; Shi Jin, Nana Liu, and Yu Yue, "Quantum Simulation of Partial Differential Equations via Schrödingerization," *Physical Review Letters*, Vol. 133, No. 23, December 6, 2024.

and vice versa,[19] so some SDEs can be solved using quantum computers just as PDEs can. Applications that have been explored include exponential calculations and computations related to mathematical finance.[20]

## The Clader-Jacobs-Sprouse Algorithm

One important application of the QLSA seems especially promising. In 2013, David Clader, Bryan Jacobs, and Chad Sprouse proposed a specific optimized version of the QLSA that could, for certain problems, provably eliminate the input/output bottlenecks that typically plague the QLSA. They showed that the specific problem of calculating RCSs satisfies all of the conditions for their optimization. Their specific optimization of the QLSA for the task of calculating RCSs is called the *Clader-Jacobs-Sprouse* (CJS) algorithm.[21]

The CJS algorithm has a much narrower range of application than the generic QLSA; it applies only to the specific problem of calculating RCSs.[22] But unlike the generic QLSA, the CJS algorithm provably gives a full exponential speedup over standard classical algorithms for solving linear systems.[23]

To summarize: The CJS algorithm is one of the rare quantum algorithms that (1) tackles a practically useful problem and (2) appears to offer an exponential speedup over the best known classical algorithm for that problem. Therefore, the CJS algorithm is arguably one of the most promising quantum algorithms known today. But the CJS algorithm has been much less well-studied than either Shor's or Lloyd's algorithms. For example, there is relatively little prior literature making quantitative resource estimates for the CJS algorithm.

With this report, we hope to help fill that gap by offering a fairly high-level assessment of the realistic prospects for the utility of the CJS algorithm. We do not attempt to do a deep dive into the algorithm itself to generate new quantitative resource estimates. Instead, we consider existing quantitative estimates and general scaling arguments to reach broad conclusions about when (if ever) quantum computers might be able to use the CJS algorithm to calculate RCSs faster than classical computers can. The main computational resource that we focus on is runtime, but we occasionally discuss memory requirements as well.

---

[19] H. Alghassi, A. Deshmuk, N. Ibrahim, N. Robles, S. Woerner, and C. Zoufal, "A Variational Quantum Algorithm for the Feynman-Kac Formula," *Quantum*, Vol. 6, June 7, 2022.

[20] Javier Gonzalez-Conde, Ángel Rodríguez-Rozas, Enrique Solano, and Mikel Sanz, "Efficient Hamiltonian Simulation for Solving Option Price Dynamics," *Physical Review Research*, Vol. 5, No. 4, December 8, 2023; Kenji Kubo, Yuya O. Nakagawa, Suguru Endo, and Shota Nagayama, "Variational Quantum Simulations of Stochastic Differential Equations," *Physical Review A*, Vol. 103, No. 5, May 21, 2021; Filipe Fontanela, Antoine Jacquier, and Mugad Oumgari, "A Quantum Algorithm for Linear PDEs Arising in Finance," *SIAM Journal on Financial Mathematics*, Vol. 12, No. 4, 2021; Elle Alhajjar, Jesse Geneson, Anupam Prakash, and Nicolas Robles, "Efficient Quantum Loading of Probability Distributions Through Feynman Propagators," arXiv, arXiv:2311.13702, version 2, November 28, 2023.

[21] Clader, Jacobs, and Sprouse, 2013.

[22] Strictly speaking, CJS proposed a more-general framework for eliminating the QLSA's input/output bottlenecks (under certain conditions) via a technique known as *preconditioning*. In principle, this preconditioned algorithm could perhaps be applied to problems other than calculating RCSs. But to our knowledge, the problem of calculating RCSs is the only problem that has been shown to meet the conditions for the CJS algorithm's preconditioning optimization. So in this report, we will use the term *CJS algorithm* to refer to the specific version of the algorithm that calculates RCSs.

[23] There is a very important caveat here. The CJS algorithm cannot solve a *generic* linear system $A\boldsymbol{x} = \boldsymbol{b}$; to our knowledge, it can solve only the specific linear system that describes RCSs. Therefore, comparing the runtimes of a quantum algorithm that specifically calculates RCSs with that of a classical algorithm that solves generic linear systems is a somewhat apples-to-oranges comparison. It is possible that there exists a narrower classical algorithm that takes advantage of the special mathematical structure of the RCS problem and matches the performance of the CJS algorithm for that specific problem. Therefore, we cannot claim that the CJS algorithm gives a *provable* exponential speedup over the fastest *possible* classical algorithm; we can say only that it gives an exponential speedup over the fastest *known* classical algorithm (Aaronson, 2015).

## Overview of the Report and Intended Audience

In Chapter 2, we define RCSs and mathematically formalize the problem of calculating them. In Chapter 3, we give a short overview of classical algorithms for calculating RCSs.[24] We show how one particular algorithm can reduce the problem to a large system of linear equations. In Chapter 4, we give a short overview of several algorithms for solving such systems of equations, including the CJS algorithm. In Chapter 5, we analyze the asymptotic (roughly speaking, qualitative) runtime of the CJS algorithm and its best classical counterpart, and we discuss in which regimes the CJS algorithm might be faster. In Chapter 6, we analyze and extend a quantitatively precise resource estimate from the previous literature. In Chapter 7, we offer some big-picture conclusions from our findings.

In this report, we use capital Roman letters ($A$) to denote matrices, boldface Roman letters ($\boldsymbol{x}$) to denote classical vectors (i.e., lists of real or complex numbers) or vector fields, and Dirac bra-ket notation ($|b\rangle$) to denote quantum state vectors. $\log()$ with no subscript denotes the natural logarithm. All vector spaces that we discuss are finite-dimensional.

This report was written by an interdisciplinary team and covers a variety of technical topics, from radar engineering to algorithmic analysis to (some) quantum hardware. We have attempted to keep it reasonably accessible to readers with expertise in any of these backgrounds, but we do not expect that many readers will be experts in all of them. To that end, we have tried to keep the main text relatively straightforward and relegated many technical details to the footnotes. None of the footnotes is critical to our main storyline. We begin a few footnotes with the phrase *technical note* to indicate that these footnotes assume specialized prior expertise and may be accessible only to specialists. Most other footnotes should be accessible for readers who want to go further into the details. We italicize technical terminology on first use; we do not expect the reader to already be familiar with those terms.

---

[24] By *classical*, we do not mean old; we simply mean that algorithms run on a classical (as opposed to a quantum) computer. Some of these algorithms were developed quite recently.

# Chapter 2

# Formalizing the Problem of Calculating Radar Cross Sections

Assume that one shines a radar pulse (or any other EM wave) on a target. Roughly speaking, an RCS captures the proportion of the incident EM wave that scatters off in various directions.

More formally: For the purpose of this report, we assume that the incident (i.e., incoming) EM wave is a simple plane wave with wave vector $\boldsymbol{k}_0$, angular frequency $\omega = c\,|\boldsymbol{k}_0|$ (where $c$ is the speed of light),[25] and amplitude vector $\boldsymbol{E}_0 \perp \boldsymbol{k}_0$.[26] We also assume that the target is made up of linear materials, which implies that it scatters EM radiation at the same frequency $\omega$ as the incident radiation.[27]

Once this incident EM wave hits a complicated target, the target will generically scatter radiation off in all directions. We denote the scattered radiation in spherical coordinates by the vector field $\boldsymbol{E}_{sc}(r, \Omega)$, where $r$ represents the distance from the scatterer and $\Omega$ the angular coordinates $(\theta, \phi)$ centered about it. Consider the outgoing direction of propagation $\boldsymbol{k}$ (with $|\boldsymbol{k}| = \omega/c$) away from the target in the angular direction $\Omega$ and some local polarization unit vector $\boldsymbol{\varepsilon} \perp \boldsymbol{k}$. Then the RCS $\sigma(\boldsymbol{k}, \boldsymbol{\varepsilon}, \boldsymbol{k}_o, \boldsymbol{\varepsilon}_0)$ roughly represents the fraction of the portion of the incident wave with initial polarization unit vector $\boldsymbol{\varepsilon}_0 \perp \boldsymbol{k}_0$ that propagates out in the $\boldsymbol{k}$ direction with polarization $\boldsymbol{\varepsilon}$:[28]

$$\sigma(\boldsymbol{k}, \boldsymbol{\varepsilon}, \boldsymbol{k}_0, \boldsymbol{\varepsilon}_0) := \lim_{r \to \infty} \frac{4\pi r^2\,|\boldsymbol{\varepsilon}^* \cdot \boldsymbol{E}_{sc}|^2}{|\boldsymbol{\varepsilon}_0^* \cdot \boldsymbol{E}_0|^2}.$$

---

[25] Strictly speaking, this is the speed of light through whatever medium the EM wave is traveling. For the earth's atmosphere, this is about 0.03-percent slower than the vacuum speed of light because of the index of refraction of air. This is a negligible correction for our purposes.

[26] Technical note: We are using the usual complex representation of the wave, whose real part represents the physical wave. So the amplitude vector $\boldsymbol{E}_0$ is a complex three-vector (sometimes called a *phasor*).

[27] Technical note: The fact that the scatterer is linear also implies that, in principle, we can calculate the scattering behavior of a more complicated pulse wavefront by Fourier superposition of the scattering amplitudes for various frequencies $\omega$. But this requires knowing the phase shift for the scattered wave of each frequency contained in the incident pulse to capture the interference effects between waves of different frequencies. These phase shifts can be challenging to measure in practice. So performing such a Fourier superposition requires knowing the complex scattering *amplitude* for each frequency $\omega$ and not just the scattering *cross-section* defined later: the absolute square of the scattering amplitude, which does not contain its phase information but is easier to measure.

[28] John David Jackson, *Classical Electrodynamics*, 3rd ed., Wiley, 1998. The case $\boldsymbol{k} = -\boldsymbol{k}_0$ is referred to as the *monostatic* cross-section because the receiver is in the same location as the source; any other case is referred to as a *bistatic* cross-section. This normalization convention is standard in radio engineering; physicists sometimes drop the factor of $4\pi$ in the numerator.

The $4\pi r^2$ compensates for the spread of the scattered field with distance from the scatterer. The complex conjugation (*) on the polarization vectors is a technical point that is not critical to understand for our purposes.[29]

Thus, the formal computational task for calculating RCSs is as follows:

- Inputs:
    - a real 3-vector $\boldsymbol{k}_0$ and a complex unit 3-vector $\boldsymbol{\varepsilon}_0 \perp \boldsymbol{k}_0$ containing the information about the incident EM wave
    - a real 3-vector $\boldsymbol{k}$ (with $|\boldsymbol{k}| = |\boldsymbol{k}_0|$) and a complex unit 3-vector $\boldsymbol{\varepsilon}$ containing the information about the scattered EM wave in the direction of interest
    - structural information about the scattering target.

- Outputs:
    - the nonnegative real number $\sigma$ defined in Equation 2.1.[30]

There are many computational algorithms for solving this problem. We discuss a few classical and quantum algorithms in Chapters 3 and 4, respectively. The nature of the input "structural information about the scattering target" depends on the precise algorithm used.

---

[29] Technical note: The polarization vectors $\boldsymbol{\varepsilon}$ and $\boldsymbol{\varepsilon}_0$ will have nonzero imaginary parts if they are basis vectors for an elliptically rather than a linearly polarized basis, so we need to complex-conjugate them when taking inner products.

[30] There are two equivalent ways to think about the information contained in the RCS function $\sigma(\boldsymbol{k}, \boldsymbol{\varepsilon}, \boldsymbol{k}_o, \boldsymbol{\varepsilon}_0)$. From a physical perspective, it is most natural to think about the RCS as inputting the information $\boldsymbol{k}_0 \in \mathbb{R}^3$ and $\boldsymbol{\varepsilon}_0 \in \mathbb{C}^3$ about the incident EM wave and outputting the intensity *profile* of the scattered radiation, which is a scalar *function* $\mathbb{R}^3 \times \mathbb{C}^3 \to \mathbb{R}^+$ of the outgoing wave parameters $\boldsymbol{k}$ and $\boldsymbol{\varepsilon}$. But from a computational perspective, it is most natural to think of the RCS function as inputting four 3-vectors $\boldsymbol{k_0}, \boldsymbol{\varepsilon_0}, \boldsymbol{k}, \boldsymbol{\varepsilon}$ and outputting a *single* scalar. These two perspectives are related via the mathematical process known as *currying*.

Chapter 3

# Algorithms for Calculating Radar Cross Sections

In this chapter, we give a short overview of several algorithmic approaches for calculating RCSs. We begin with a high-level overview of the many different classical approaches to this problem. We then delve in more detail into one particular approach—the *frequency-domain finite-element method* (FD-FEM)—which is the foundation of the CJS algorithm, discussed in the next chapter.[31]

## Overview of Classical Approaches

To talk about classical approaches, we will first briefly discuss what they need to capture. In this section, we will introduce a brief review of the various *scattering mechanics* or the ways in which electromagnetic energy is absorbed, reflected, and refracted. We will then briefly discuss the three dominant *regimes* in which varying assumptions can be made about which types of scattering mechanisms dominate because that influences the choice over which scattering mechanisms are most critical to include in an analysis.[32]

Once we have reviewed those two basic facets of scattering, we will introduce a taxonomy of how the EM effects may be calculated, including the choice between simulation and measurement, and several key simulation approaches.

After this section, we will move on to a more detailed review of the FD-FEM because that is the algorithm most closely related to the CJS quantum algorithm that we are investigating.

### Scattering Mechanics

To understand the various computational methods, we must first discuss the methods by which EM energy is scattered and reflected. There are many physical properties of an object that contribute to its RCS, and the mechanics of how each contributes are complex and codependent, but we can list several key factors.[33]

---

[31] We do not mean to imply that the FD-FEM is the best or most widely used classical approach used for calculating RCSs today. It is simply the only approach that is known to admit a dramatic quantum speedup over classical algorithms.

[32] Most of the material in this section, including both the discussion of scattering mechanics and the methods for approximating a target's signature, is adapted from Chapter 14 of Merrill I. Skolnik, ed., *Radar Handbook*, 3rd ed., McGraw Hill, 2009.

[33] Skolnik, 2009, Chapter 14.

- **Specular returns:** Signals are reflected off surfaces according to Snell's Law (the reflection angle and incident angle have the same angle relative to the surface at the point of reflection).
- **Tip, corner, and edge diffraction:** Signals are diffracted and scattered by sharp tips, corners, or edges.
- **Gap and seam echoes:** Discontinuities in the surface generate diffraction return.
- **Curvature discontinuities:** Changes in the radius of curvature of an object cause reflections.
- **Creeping wave returns:** Incident EM fields induce a current in the object, which travels around the skin and regenerates a new EM field when it circumnavigates the object.
- **Traveling wave echo:** The induced current will reflect off internal structures and generate new EM fields.
- **Cavity returns:** EM waves entering a cavity will undergo many reflections before eventually exiting.

Typically, the choice of which features to include in a computational model is a balance between runtime and fidelity. Many of the scattering mechanisms will have limited impact on the RCS for various objects, depending on geometry (how complex is the object's shape), material properties (e.g., conductance and resistance), and spectral regime (wavelength at which the RCS is calculated relative to the size of an object's features). Thus, EM modelers typically employ a variety of tools and use their expertise to determine which is most applicable in a given scenario.

## Spectral Regions

A key factor in the RCS of an object is the relationship between the object's size (or the size of its characteristic features) and that of the incident signal's wavelength. At lower frequencies, when the wavelength is much larger than the object, the interaction is characterized by the *Rayleigh regime*, wherein the principal effect is the coupling of the incident field to currents in the surface of the object; the object then reradiates new EM fields. At the other extreme, where the size of the object is large compared with the wavelength, optical theory dominates, and the interaction is well characterized by treating the EM field as a series of rays bouncing off the objects; this is the *optical regime*. Between the two is a *resonance regime* in which the dominance of one effect over the other is not guaranteed.

## Taxonomy of Calculation Methods

There are many ways to calculate an object's RCS. The choice of which approach to use varies with parameters, such as the frequency of interest, and physical realities, such as the presence of a physical prototype.

### Measurement Versus Simulation

The most reliable method for calculating an RCS is to measure it experimentally. The object under test is placed on a pylon, which is typically covered in radar absorbing panels, and the device is scanned repeatedly by probing it with either single-frequency tones or wideband impulse signals and recording the reflections. The object is then rotated and the process repeated. The principal drawback of measurement approaches is the time required to collect precise samples at the required orientations and frequencies. The other is the need for a physical object to scan. There are countless variations on this theme, including the use of subscale models to stand in for larger test objects, waveform choices, and the choice of test ranges.

The alternative to measurement is to calculate the RCS using one of the many simulation approaches available, which we broadly categorize into exact and approximate methods. Despite their names, both methods are inexact, and the level of precision required will dictate computational requirements.

Exact Methods

The exact methods for RCS calculation (or, more precisely, for EM modeling) are referred to this way because they rely on direct calculation of Maxwell's equations.

The first exact method is the *method of moments*. In this formulation, Maxwell's equations are solved in integral form to compute exact currents along the surface of an object in response to a probing signal. To do this computationally, the surface of the objects is discretized as a mesh. The complexity of this method is $O(L^M N^M)$ for an object's length $L$, the number of spatial dimensions $M$, and the number of mesh points $N$ used to discretize the object's surface. Clever methods, such as taking advantage of symmetry, can be used to simplify the problem.

The second method is two paired approaches: *finite-difference time domain* (FD-TD) and *finite-difference frequency domain* (FD-FD) methods. Their uses are often complementary. FD-FD is much faster but generates only a single-frequency result at a time and does not consider transient effects. FD-TD is more time-consuming but generates a time-dependent response that can capture transients and can be used to analyze multiple frequencies at once.

In both cases, the volume around the object in question is discretized. At each point, Maxwell's equations are reformulated as difference equations allowing the exact calculation of the EM fields at a given point in space that is based on the fields of neighboring points in the discretized grid. In FD-TD, the input is a time domain signal, and the difference equations are used to simulate its progression across space, interactions with the object under test, and reflection back to the point of origin. FD-FD is a slightly different formulation that gives only the steady state response and generates only single-frequency results (to analyze multiple frequencies, repeated runs are required). Complexity scales with $O(L^M N^M)$ for the simulation area length $L$, number of spatial dimensions $M$, and number of grid points $N$ per spatial dimension for each frequency sample.

Another class of discretized solution is FEMs. Where FD methods discretize space into a series of points at which to compute fields and interactions, FEMs discretize space into volumes and solves for the EM fields within each volume. As with FD, FEM can be computed in time or frequency domains. Because FEM discretizes the scene into areas or volumes instead of points, the choice of discretization is different, and consideration is typically given to the shape and packing of various elements, such as pyramids or triangles as opposed to the squares or cubes that are commonly used in FD methods. We will discuss the FD-FEM technique in more detail in the following section.

Approximate Methods

There are many approximate methods; the primary benefit of each is the vastly reduced computational runtime when compared with the exact methods. These methods are most accurate in high-frequency problems when scattering is well into the optical region. We will not discuss them in detail, but an abbreviated list of the popular techniques is as follows:[34]

- **Geometric optics (GO):** Perform simple ray tracing; discretize the object as a mesh and treat each facet as doubly-curved with some radius $R$. GO accounts for refraction but not specular returns. GO is invalid in shadows and does not consider polarization.
- **Physical optics (PO):** Use ray optics to estimate the field on a surface and integrate the field over the surface to calculate the transmitted or scattered field. PO can support flat or curved facets but does not consider polarization.
- **Geometric theory of diffraction (GTD):** This is an improvement of GO that adds support for edge responses and handling of tip and corner diffraction.
- **Physical theory of diffraction (PTD):** This is an improvement of PO that corrects for known errors in the surface current calculation that increases accuracy near sharp edges and corners.

In general, approximate methods are characterized by a balance of the desired level of fidelity and acceptable computational impact. They can be a great companion to exact methods because their more rapid calculation allows for iterative checks as a design is improved, but they are not considered authoritative.

## Calculation of Radar Cross Sections

All the aforementioned methods involve explicitly modeling or measuring the scattered response to a probing signal. Calculation of the RCS is a straightforward process of dividing the response signal by the probing signal after accounting for the various losses (including propagation loss) in the system.

## The Frequency-Domain Finite-Element Method

In this section, we go into more detail about the FD-FEM, which is the approach that the CJS algorithm uses as its starting point.[35]

In SI units and standard notation, Maxwell's equations in a linear, isotropic, homogeneous medium (which we can approximate air to be) are as follows:

$$\nabla \cdot \boldsymbol{E} = \frac{\rho}{\epsilon},$$
$$\nabla \cdot \boldsymbol{B} = 0,$$

---

[34] Skolnik, 2009, Chapter 14.

[35] The FEM is the only approach to calculating RCSs that is known to admit a quantum speedup. Therefore, for problems for which approximate methods are more powerful than exact methods, a quantum speedup for the FEM may or may not prove useful in practice. Of course, if the quantum speedup is very significant, then the quantum FEM might supplant classical approximate methods for certain problems.

$$\nabla \times \boldsymbol{E} = -\frac{\partial \boldsymbol{B}}{\partial t},$$
$$\nabla \times \boldsymbol{B} = \mu \boldsymbol{J} + \mu\epsilon \frac{\partial \boldsymbol{E}}{\partial t}.$$

Air typically has negligible charge and current density, so we can set $\rho = 0$ and $\boldsymbol{J} = \boldsymbol{0}$ and consider the vacuum equations (subject to certain boundary conditions that we discuss later).[36] We can take the curl of the last two equations to decouple them to the following vector wave equations:

$$\nabla^2 \boldsymbol{E} - \mu\epsilon \frac{\partial^2 \boldsymbol{E}}{\partial t^2} = \boldsymbol{0},$$
$$\nabla^2 \boldsymbol{B} - \mu\epsilon \frac{\partial^2 \boldsymbol{B}}{\partial t^2} = \boldsymbol{0}.$$

We assume that the scattering target is made of linear materials and that its structural properties (e.g., size, shape, and absorption properties) do not change significantly over time. Then the system is a linear time-invariant system, and we can Fourier transform these equations to the frequency domain and use the relation $k = \sqrt{\epsilon\mu}\,\omega$ to get the following time-independent equations:

$$(\nabla^2 + k^2)\,\boldsymbol{E} = \boldsymbol{0},$$
$$(\nabla^2 + k^2)\,\boldsymbol{B} = \boldsymbol{0}.$$

So we have reduced Maxwell's equations to two decoupled copies of the (complex) vector Helmholtz equation, and we have reduced the domain from four-dimensional spacetime to three-dimensional space.[37] We will focus on the first equation,

$$(\nabla^2 + k^2)\,\boldsymbol{E} = \boldsymbol{0}, \tag{3.1}$$

for the electric field because that is the one conventionally used for calculating RCSs.

There are many different methods to tackle the three-dimensional vector Helmholtz equation with complicated boundary conditions, such as a complicated scattering target, including the FD or FEMs discussed previously for solving PDEs.[38]

The CJS algorithm uses the FEM. The FEM is quite complicated, and the details are not critical for the purpose of this report, so we will give only a flavor of how it works.[39] The idea behind the FEM is that we divide space up into a mesh of small polyhedral volumes known as *finite elements*.[40] This mesh consists of a large number of point *nodes* at the shared corners of adjacent finite elements, and adjacent nodes are joined by straight *edges*.[41] Each finite element contains a small number of degrees of freedom, and the differential operators in the PDE are translated into algebraic relations between the degrees of freedom in neighboring finite elements. The resulting system of algebraic equations is linear if (and only if) the PDE is linear.

---

[36] For simplicity, Clader, Jacobs, and Sprouse (2013) assumed that the scatterer was a perfect conductor and did not model the propagation of the EM waves inside the scatterer itself. But the CJS algorithm can also be generalized to handle non-vacuum solutions and model the propagation within the scatterer. Doing so would increase the computational requirements discussed later but should not change their qualitative scaling patterns.

[37] There is no free lunch here: We need to separately solve Equation 3.1 for many different values of $k$ (or equivalently, frequencies $\omega$) to reconstruct the scattering of a wideband incident pulse.

[38] Jian-Ming Jin, *The Finite Element Method in Electromagnetics*, 2nd ed., Wiley, 2002.

[39] Jin, 2002.

The CJS algorithm uses a specific version of the FEM known as the *edge-based FEM*.[42] In the traditional formulation of the FEM, the field values are defined on the nodes of the mesh, but in the edge-based FEM, the field values are instead defined on the mesh's edges. Each edge $e_i$ is given an orientation. The degrees of freedom are the components of the electric field $\boldsymbol{E} \cdot \boldsymbol{e}_i$ at the location of the edge $\boldsymbol{e}_i$. Therefore, there is a single scalar degree of freedom along each edge of the mesh.[43] As the mesh becomes very dense, we get a high-resolution discretization of the electric field.

The edge-based FEM is better suited than the node-based FEM for dealing with scattering geometries with sharp edges or corners, where the electromagnetic fields become singular. It also naturally captures the boundary conditions at an interface between two different materials (namely, continuity of the tangential component of the electric field and discontinuity of the normal component).[44] Moreover, the edge-based FEM can be implemented in a way that automatically guarantees that the electric field will be divergence-free, as Gauss's law requires.

Each finite element generates a set of linear equations relating the degrees of freedom at each of that element's edges. So, for example, for a uniform square mesh, each finite element (a square in the mesh) generates an equation relating four variables that correspond to the four edges of the square. For a uniform triangular mesh, each finite element (a triangle) generates an equation relating the three degrees of freedom that live on the triangle's three edges.

The result of the FEM is to convert the Helmholtz equation into a large system of $N$ linear equations $A\boldsymbol{x} = \boldsymbol{b}$, where $N$ is the number of edges in the mesh, $A$ is a known $N \times N$ matrix, $\boldsymbol{b}$ is a known $N$-component vector, and $\boldsymbol{x}$ is an unknown $N$-component vector to be solved for. The matrix $A$ encodes the Helmholtz equation, the geometry of the mesh, and the geometry of the scattering target. The vector $\boldsymbol{b}$ encodes the known electric fields at the scattering target's boundary. The unknown vector $\boldsymbol{x}$ encodes the electric field at each edge of the mesh. The exact formulas for $A$ and $\boldsymbol{b}$ are very complicated and not necessary for the purpose of this report.

The vector $\boldsymbol{x}$ gives a discretized description of the full electric field $\boldsymbol{E}$ over all of the modeled space. The actual radar cross section is

$$\sigma = \frac{1}{4\pi} |\mathbf{R} \cdot \mathbf{x}|^2, \tag{3.2}$$

where $\mathbf{R}$ is an $N$-component vector that depends on the direction $\mathbf{k}$ and the polarization $\boldsymbol{\varepsilon}$. The exact formula for $\mathbf{R}$ involves a complicated surface integral,[45] but $\mathbf{R}$ can be efficiently computed numerically.

---

[40] In general, the FEM can use an arbitrary mesh with different sizes and shapes of finite elements in different places. In many cases, the mesh is deliberately chosen to be irregular and has a higher density of finite elements in places where the field changes quickly or where higher resolution is required. But as we will discuss later, the CJS algorithm requires a regular, periodic mesh.

[41] To clarify some potentially confusing terminology: The *edges* of the mesh occur throughout the bulk of the mesh. We will use the term *boundary* of the mesh to refer to its outer surrounding surface where the mesh terminates.

[42] A. Chatterjee, J. M. Jin, and J. L. Volakis, "Edge-Based Finite Elements and Vector ABCs Applied to 3-D Scattering," *IEEE Transactions on Antennas and Propagation*, Vol. 41, No. 2, February 1993.

[43] Technical note: Clader, Jacobs, and Sprouse (2013) list the specific edge basis functions that they use in their edge-based FEM in Equations 65 and 66 of their supplementary material.

[44] Chatterjee, Jin, and Volakis, 1993.

A very important feature of the FEM is that it generates a *sparse* system of linear equations. Loosely speaking, this means that almost all of the entries of $A$ equal zero. More formally, the maximum number of nonzero elements in each row of the matrix $A$ is independent of the total size of the matrix.[46] Later in this report, we will return to the important question of just *how* sparse the matrix is.

---

[45] Clader, Jacobs, and Sprouse, 2013, Equation 15.

[46] Some sources define the term *sparse* more broadly to allow the number of nonzero entries per row to grow polylogarithmically in the size of the matrix. But in this report, we will consider only matrices with a constant number of nonzero entries per row.

Chapter 4

# Algorithms for Solving the Linear System of Equations

In Chapter 3, we formulated Maxwell's equations in the frequency domain and reduced them to the Helmholtz equation. We then showed how we can use the FEM to reduce the Helmholtz equation to a large but sparse system of linear equations of the form $Ax = b$. This setup is the same for both the classical and quantum approaches; the two approaches diverge in how to solve this linear system for $x$ (and therefore for the electric field).

## Classical Algorithm for Solving the System

The standard classical algorithm for solving this sparse linear system is the *conjugate gradient* (CG) method.[47] We will not discuss the details of this method, but it is a very standard and well-understood algorithm. The only information that we will need about this algorithm is its asymptotic runtime, which we will discuss in Chapter 5.

## Quantum Algorithm for Solving the System

### Quantum Linear System Algorithms

We first step back and discuss, in general, how one would solve a linear system on a quantum computer. We now consider a general linear system $Ax = b$, which does not necessarily have anything to do with RCSs. There are several quantum algorithms for efficiently solving this system for $x$. These algorithms are all variations on the same general idea, which is known as the QLSA, an umbrella term that covers all these algorithms.

The original version of the QLSA was the 2008 HHL algorithm.[48] In 2010, Andris Ambainis discovered a new variation of the QLSA that slightly improved on the HHL algorithm by improving its runtime's dependence on the condition number of the matrix.[49]

In the setup for the QLSA, we assume that $A$ is an $N \times N$ complex matrix and $b$ is a complex $N$-component vector. We will refer to $b$ as a *classical vector*, meaning that it is simply a list of numbers. We encode the classical vector $b$ in the amplitudes of a quantum state vector $|b\rangle$ on $n \approx \log_2 N$ qubits. Similarly, we encode the matrix $A$ as a quantum operator $\hat{A}$ acting on the $n$ qubits.[50] The output of the QLSA is a quantum state vector $|x\rangle$ whose amplitudes encode the classical solution vector $\boldsymbol{x}$. We then need to read out those amplitudes, either by directly measuring the state vector $|x\rangle$ (which is a probabilistic process in quantum mechanics and so gives only statistical information about $|x\rangle$) or by performing further quantum operations before measurement.

---

[47] Yousef Saad, *Iterative Methods for Sparse Linear Systems*, Society for Industrial and Applied Mathematics, 2003.

[48] Harrow, Hassidim, and Lloyd, 2009.

[49] Andris Ambainis, "Variable Time Amplitude Amplification and a Faster Quantum Algorithm for Solving Systems of Linear Equations," arXiv, arXiv:1010.4458, November 14, 2010. We will explain the term *condition number* in the next section.

Remarkably, for certain matrices $A$, the QLSA can compute the quantum state vector $|x\rangle$ exponentially faster than the best known classical algorithm can calculate the corresponding classical vector $\boldsymbol{x}$.[51] The fact that we can encode $N$ equations into only $\log_2 N$ qubits is the key feature that allows for this exponential speedup.

Figure 4.1 illustrates the overall process flow.

**Figure 4.1. Logical Flow of the Quantum Linear System Algorithm**

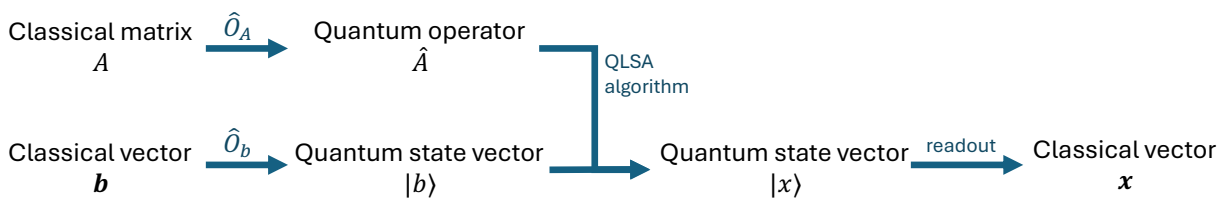

We see that the overall logic flow consists of four transformations:

1. encoding the classical vector $\boldsymbol{b}$ into the quantum vector $|b\rangle$
2. encoding the classical matrix $A$ into a quantum operator $\hat{A}$
3. using the quantum operator $\hat{A}$ to convert the quantum vector $|b\rangle$ to a quantum vector $|x\rangle$
4. reading the quantum vector $|x\rangle$ out into the classical vector $\boldsymbol{x}$ (or some derived quantity).

None of these four steps is easy. Most QLSA algorithms focus only on step 3; they assume that $\hat{A}$ and $|b\rangle$ are already given as quantum-encoded inputs, and they conclude by returning the quantum state $|x\rangle$ as an output. But this process is not a full end-to-end algorithm; it can serve only as a

---

[50] The state of a quantum computer is represented by a vector in a very large vector space called the *Hilbert space*. If the quantum computer contains $n$ qubits, then the dimension of the Hilbert space is $N = 2^n$. Therefore, it takes an exponentially large amount of information to specify the state of $n$ qubits—or conversely, $n$ qubits can store an exponentially large amount of information (although there are some important nuances around the word *store*). The notation $|\ \rangle$ denotes a quantum state vector in this Hilbert space. A hat symbol ˆ over a letter denotes a linear operator acting on this Hilbert space. A quantum algorithm consists of a sequence of (unitary) operators applied to an initial state vector, followed by a final measurement of the quantum state.

Technical note: the quantum operator $\hat{A}$ usually needs to be Hermitian for the QLSA to work. So we sometimes need to perform some technical tricks to convert the classical matrix $A$ to a Hermitian operator $\hat{A}$, which sometimes requires adding in some additional qubits.

[51] Technical note: Harrow, Hassidim, and Lloyd (2009) demonstrated that certain complexity-theoretic arguments very strongly suggest that the HHL algorithm is exponentially faster than the fastest *possible* classical algorithm for a *generic* matrix $A$ specified by a black-box oracle. As we will discuss later, the precise requirements for the exponential speedup are that the matrix $A$ be sparse, well-conditioned, and efficiently computable. The first two requirements mean that the number of nonzero entries per row of $A$ and the condition number of $A$ must scale at most polylogarithmically in $N$.

subroutine of a larger algorithm that must begin with the classical vector **b** and end with the classical vector **x**.[52]

More precisely, the QLSA usually assumes that the quantum vector $|b\rangle$ is initialized by some *oracle*—a black-box quantum (unitary) operator $\hat{O}_b$ whose internal details are unspecified such that $|b\rangle = \hat{O}_b|0\rangle$. Similarly, the QLSA assumes that it is known which (few) entries of the sparse matrix $A$ are nonzero. The quantum operator $\hat{A}$ is typically instantiated by another oracle, the *matrix oracle*,[53] a quantum operator that inputs two indices $i, j = 1, \ldots, N$ and outputs the matrix element $A_{ij}$. Under these assumptions, the QLSA provides an exponential speedup *for step 3* compared with a classical computer.[54]

Most discussions of the QLSA simply assume these oracles as given, but any actual implementation of QLSA on a quantum computer must implement both of these oracles with specific quantum circuits. If these circuits cannot be implemented efficiently, then they often act as bottlenecks that neutralize any speedup from the rest of the algorithm.[55] The oracle formulation is a useful abstraction, but it can sometimes hide these bottlenecks and exaggerate the magnitude of the quantum speedup.

To understand the full speedup of an actual application of the QLSA to a specific problem, we need to consider the end-to-end runtime of the entire algorithm, not just of step 3. Even if the QLSA hugely speeds up step 3, it does not necessarily speed up the entire calculation if that speedup simply shifts the bottleneck to step 1, 2, or 4. The generic QLSA does not give an end-to-end exponential speedup for most real-world problems, where the inputs and outputs are generic unstructured classical vectors and input/output bottlenecks neutralize the speedup of step 3.[56]

## The Clader-Jacobs-Sprouse Algorithm

Clader, Jacobs, and Sprouse instead found an efficient quantum algorithm for solving the same linear system for the case of calculating RCSs. Their discovery consisted of two major pieces that we will briefly summarize: (1) a *preconditioning* step that speeds up the computation and (2) explicit constructions of efficient quantum oracles both initializing the vector $|b\rangle$ and implementing the matrix $\hat{A}$.

The preconditioning step is quite technical, so we will discuss it only briefly. Effectively, it greatly reduces the *condition number* of the matrix $A$. The condition number of a matrix summarizes how numerically sensitive it is to perturbations; the higher the condition number, the more difficult it is to solve the system accurately.[57] The runtime of the original HHL algorithm depends sensitively on the condition number, and for a generic matrix $A$, the HHL algorithm is too slow to deliver an exponential speedup over classical methods.

---

[52] Or some derived quantity, such as the inner product $\boldsymbol{x} \cdot \boldsymbol{R}$ for a given vector $\boldsymbol{R}$.

[53] Not to be confused with the character of The Oracle from the movie *The Matrix*.

[54] More precisely, the QLSA provides an exponential reduction in the number of calls to the matrix oracle relative to a classical algorithm.

[55] Aaronson, 2015.

[56] Aaronson, 2015.

[57] Technical note: The condition number $\kappa$ of a matrix $A$ is defined to be the maximum ratio of the relative error in the vector $\boldsymbol{x}$ to the relative error in $\boldsymbol{b} = A\boldsymbol{x}$ over all vectors $\boldsymbol{x}$. For an invertible matrix $A$, the condition number equals the product

The key idea behind the CJS algorithm's preconditioning step is that instead of directly solving the equation $A\boldsymbol{x} = \boldsymbol{b}$, it instead solves the related equation $MA\boldsymbol{x} = M\boldsymbol{b}$, where $M$ is a new matrix. Clader, Jacobs, and Sprouse constructed the matrix $M$ so that the product $MA$ would remain sparse and would have a much smaller condition number than $A$, which allowed them to then apply the HHL algorithm much more efficiently than before. This preconditioning trick does not always work for a general linear system, but they showed that it does work for the specific linear system that arises from the RCS problem.[58]

Clader, Jacobs, and Sprouse's second major discovery was arguably even more important: They gave explicit constructions for the oracles for both the quantum-encoded vector $|b\rangle$ and the quantum operator $\hat{A}$ for the case of the RCS problem, and they showed that both constructions are efficient.[59] Therefore, they demonstrated an *end-to-end* algorithmic speedup over classical methods that covered all four aforementioned steps with no bottlenecks. The CJS algorithm for calculating RCSs is one of very few known *end-to-end* quantum speedups for solving linear systems.[60]

## Implementation Details

This subsection goes into more detail about how Clader, Jacobs, and Sprouse converted the RCS problem into a system of linear equations. It is not critical for later sections in this report.

For simplicity, they considered a two-dimensional metallic scatterer, although their techniques can be generalized to handle more-complicated boundary conditions. They considered two-dimensional space, which they discretized into a uniform square mesh. They modeled the scatterer as a perfect conductor, which requires that the electric field at the scattering surface be perpendicular to it.[61] Therefore, at the scattering surface, $\hat{n} \times \boldsymbol{E}_{total} = \hat{n} \times (\boldsymbol{E} + \boldsymbol{E}^i) = \boldsymbol{0}$, where $\hat{n}$ is the unit vector normal to the scattering surface and $\boldsymbol{E}^i$ and $\boldsymbol{E}$ are the incident and scattered fields, respectively. This

---

$\|A^{-1}\| \|A\|$, where $\|A\|$ represents the operator norm of $A$. If $A$ is Hermitian, as is the case for the CJS algorithm, then $\|A\|$ is its largest eigenvalue (in absolute value), so the condition number equals the ratio of (the absolute values of) its largest eigenvalue to its smallest eigenvalue.

[58] Technical note: The precise requirement for the CJS algorithm's trick to work is that the matrix must admit a *sparse approximate inverse* preconditioner. Marcus J. Grote and Thomas Huckle, "Parallel Preconditioning with Sparse Approximate Inverses," *SIAM Journal on Scientific Computing*, Vol. 18, No. 3, 1997; Edmond Cho, "A Priori Sparsity Patterns for Parallel Sparse Approximate Inverse Preconditioners," *SIAM Journal on Scientific Computing*, Vol. 21, No. 5, 2000.

[59] More precisely, the algorithm's runtime scales only logarithmically in $N$, the number of equations in the system. Recall that the CJS algorithm also requires the computation of a vector $\boldsymbol{R}$ in Equation 3.2 to extract the final RCS. Constructing the corresponding quantum state $|R\rangle$ requires a third oracle, whose resource costs are very similar to those of the state preparation oracle $\hat{O}_b$.

[60] Aaronson, 2015.

[61] This assumption is not necessary; the CJS algorithm can be straightforwardly generalized to handle more-realistic scatterers with imperfect conductivity, positive skin depth, and so on. The CJS algorithm's authors chose perfect-conductor boundary conditions simply to illustrate the general algorithm. An interesting direction for future research would be to explore how a more realistic and detailed representation of the scatterer would increase the complexity (and therefore the computational resource requirements) of the oracles containing the structural information about the scatterer.

boundary condition imposes the requirement that the tangential components of the incident and scattered fields be negatives of each other.[62]

As we discussed earlier, one important requirement for the CJS approach to work is that elements of the matrix $A$ must be efficiently calculable. Recall that, in general, the FEM can be used on an arbitrary mesh; indeed, this is one of the main advantages of the method. But the matrix $A$ is complicated to calculate on an arbitrary mesh. The requirement that $A$ be efficiently calculable essentially requires the mesh to be regular or semiregular (e.g., a square, triangular, or cubic lattice) for the CJS algorithm to be efficient.[63] This is an important limitation on the CJS algorithm.

### Another Promising Possibility: The Childs-Kothari-Somma Algorithm

At the risk of overloading the reader with yet another hard-to-remember algorithm made up of three people's names, we briefly discuss another quantum algorithm that may prove promising: the Childs-Kothari-Somma (CKS) algorithm, discovered in 2017.[64] Similar to the HHL algorithm, the CKS algorithm is a quantum algorithm for solving large systems of linear equations. But it works on fundamentally different principles from the HHL algorithm, and it has certain theoretical advantages over HHL, most notably an *exponentially* better dependence on the error tolerance $\epsilon$.[65]

The CKS algorithm could potentially be applied to the same linear system $A\boldsymbol{x} = \boldsymbol{b}$ for the RCS problem that the CJS algorithm solves. If so, it could potentially offer a qualitative improvement over the CJS algorithm by greatly reducing the resource costs that we will discuss in the next two chapters.

But there are some complicated technical requirements for when the CKS algorithm can be used.[66] A full analysis of whether the CKS algorithm can be inserted into the CJS algorithm (without losing the CJS algorithm's exponential speedup) is beyond the scope of this report. We note only that this possibility appears promising, but, as we will briefly discuss in Chapter 7, even replacing the HHL subroutine within the CJS algorithm with a faster CKS subroutine would not necessarily deliver a full end-to-end exponential speedup of the complete CJS algorithm. Nevertheless, exploring this avenue is an excellent opportunity for further research.

## Summary

The advantage of the CJS approach is that, in some cases, it can be much, much faster than the classical CG algorithm. We will consider its runtime in detail in the following two chapters, but the key point is that the runtime of the classical CG method scales as $N$, while the runtime of the CJS algorithm only scales as $\log(N)$, so the CJS algorithm is *exponentially* faster than the classical CG algorithm—the gold standard for quantum speedups.

---

[62] Technical note: At the outer boundary surface of the entire space, the authors also imposed absorbing boundary conditions to suppress reflections off the artificial boundary of the region being modeled.

[63] Clader, Jacobs, and Sprouse, 2013, supplementary material.

[64] Andrew M. Childs, Robin Kothari, and Rolando D. Somma, "Quantum Algorithm for Linear Systems of Equations with Exponentially Improved Dependence on Precision," *SIAM Journal on Computing*, Vol. 46, No. 6, 2017.

[65] The CKS algorithm also has some more-technical advantages over the HHL algorithm, such as avoiding the need for postselection.

[66] Technical note: For example, the matrix $A$ must be efficiently block-encodable.

However, as we mentioned in the introduction, the CJS algorithm has not been *proven* to be faster than the fastest possible classical algorithm for calculating RCSs. It has not even been proven to be faster than the fastest possible classical algorithm for calculating RCSs using the frequency-domain FEM. It has been proven only to be faster than the CG algorithm, which is the best *currently known* classical algorithm for calculating RCSs using the frequency-domain FEM. The CG method is a generic algorithm for solving sparse linear systems, and this algorithm is not specialized to the problem of calculating RCSs as the CJS algorithm is. In other words, we are comparing a special-purpose quantum algorithm with a general-purpose classical algorithm.

Someone could potentially discover a corresponding special-purpose classical algorithm for efficiently solving the particular linear system $A\boldsymbol{x} = \boldsymbol{b}$ that emerges from the RCS problem, which takes advantage of the particular structure of Maxwell's equations to match the performance of the CJS algorithm.[67] As is usually the case with quantum computing, the apparently quantum speedup is only provisional, and it could be eliminated at any time if someone finds a classical algorithm that matches its performance.

---

[67] Aaronson, 2015.

Chapter 5

# Asymptotic Runtime Analysis of the Clader-Jacobs-Sprouse Algorithm

In this chapter and in Chapter 6, we discuss the computational resource and time costs associated with running the CJS algorithm on a hypothetical quantum computer. In this chapter, we perform an *asymptotic analysis*; that is, we consider only the qualitative form of the functional dependence of the runtime on the problem size $N$ in the limit of large $N$ without worrying about making precise quantitative predictions (e.g., calculating overall multiplicative constants). This rough analysis is very common in studying quantum algorithms. In the next chapter, we will consider quantitatively precise estimates.

## Asymptotic Runtimes of the Clader-Jacobs-Sprouse and Conjugate Gradient Algorithms

### Variable Definitions

$A\boldsymbol{x} = \boldsymbol{b}$ represents the linear system to be solved. $T_{CG}$ and $T_{CJS}$ represent the runtimes of the classical CG algorithm and the quantum CJS algorithm, respectively, as measured in abstract logical time steps.[68] The positive real number $\kappa$ represents the condition number of the matrix $A$. The positive integer $d$ represents the *sparsity* of the matrix $A$, the maximum number of nonzero entries that appear in any row of $A$. $\epsilon \in [0,1]$ represents the precision of the solution: the relative error in the computed vector $\boldsymbol{x}$ relative to its exact value.[69] The appendix gives some quantitative intuition for which precisions of RCSs are sufficient for which practical tasks.

$N$ represents the number of equations in the linear system $A\boldsymbol{x} = \boldsymbol{b}$; $A$ is an $N \times N$ complex matrix and $\boldsymbol{b}$ is an $N$-component complex vector. Physically, $N$ represents the number of edges in the finite-element mesh into which the scattering space is discretized. This is the main dependent variable whose asymptotic scaling we will be considering.

---

[68] The exact definition of a *logical time step* is not critical in this chapter, because we will not yet be considering numerical constants. For concreteness, the reader can think of a logical time step as representing one floating- or fixed-point arithmetic operation or one processor clock cycle.

[69] More precisely, $\epsilon$ is the multiplicative error in the final RCS $\sigma \propto |\boldsymbol{R} \cdot \boldsymbol{x}|^2$ for a given mesh discretization. That is,
$$\sigma(1-\epsilon) \leq \tilde{\sigma} \leq \sigma(1+\epsilon),$$
where $\sigma$ is computed from the exact solution $\boldsymbol{x}$ to the linear system and $\tilde{\sigma}$ is computed from the estimated solution. As we discuss in the appendix, the exact RCS also has an additional error contribution from the discretization itself.

$C_{CG}$ and $C_{CJS}$ are constant prefactors that capture the number of logical operations performed in the classical or quantum algorithms, respectively. Asymptotic runtime analysis does not attempt to quantify the values of these constant prefactors, which is very hard to do; we will not try to estimate these values until the next chapter.

## Asymptotic Runtimes

The classical CG algorithm has asymptotic runtime

$$T_{CG} \sim C_{CG}\, \kappa\, d\, \log(1/\epsilon)\, N \tag{5.1}$$

as $N \to \infty$, while the quantum CJS algorithm has asymptotic runtime

$$T_{CJS} \sim C_{CJS}\, \kappa\, d^7\, \epsilon^{-2}\, \log N \tag{5.2}$$

as $N \to \infty$.[70]

Let us begin by qualitatively considering the growth with respect to each parameter:

- Both algorithms scale the same way in the condition number, as $\kappa^1$. Therefore, neither algorithm has a *relative* advantage with respect to the condition number; we do not expect the condition number to affect which algorithm performs better.
- The CJS algorithm scales much worse than the classical CG algorithm in the sparsity: $d^7$ for the CJS algorithm versus only $d^1$ for CG. This implies that the sparsity is *very* important in determining the relative advantage: Unless the system is *very* sparse (e.g., $d = 2$ or $d = 3$), the CJS algorithm will be at a significant disadvantage to the CG algorithm.[71]
- The CJS algorithm also scales much (exponentially) worse than the CG algorithm in the precision: $\epsilon^{-2}$ for CJS versus only $\log(1/\epsilon)$ for CG.
- But the CJS algorithm scales much (exponentially) better than the CG algorithm in the problem size: only $\log N$ for the CJS algorithm versus $N$ for the CG algorithm.

---

[70] Clader, Jacobs, and Sprouse, 2013. Without the preconditioning step of the CJS algorithm discussed in Chapter 4, $\kappa$ depends implicitly on $N$ as $\kappa = O(N^{M/2})$, where $M$ is the number of spatial dimensions. So, without the preconditioning, the CJS algorithm's runtime's implicit dependence on $N$ via $\kappa$ would actually dominate its explicit logarithmic dependence on $N$ in Equation 5.2, eliminating the exponential speedup. But the preconditioning step bounds the value of $\kappa$ to a constant that does not depend on $N$, thereby delivering an exponential speedup.
Technical note: Clader, Jacobs, and Sprouse used the $O()$ notation in their expression for Equation 5.1 but the $\tilde{O}()$ notation in their expression for Equation 5.2, indicating that they have dropped more–slowly growing multiplicative corrections. These slowly growing corrections come from an omitted factor of

$$\exp[2\sqrt{\ln(5)\ln(d^2\kappa/\epsilon^2)}],$$

which grows faster than polylogarithmically in $d^2\kappa/\epsilon^2$ but slower than any power law.

[71] Technical note: We can decompose the $d^7$ dependence into a product of ($d^3$ time steps to apply the preconditioner $M$) × ($d^4$ exponential operator applications in the Hamiltonian simulation), where we have dropped the dependence on all other parameters. This upper bound of $d^7$ is not necessarily tight; improved implementations of the CJS algorithm might be able to decrease the very strong runtime dependence on $d$ to an asymptotically slower growth rate, which would hugely speed up the algorithm.

We see that, all else being equal, the classical CG algorithm tends to perform better for linear systems that are only somewhat sparse and have a moderate number of equations but where high precision is very important. By contrast, the quantum CJS algorithm tends to be perform better for systems that are *very* sparse and contain a huge number of equations but that only require moderate precision for the solution.

These facts suggest that if the CJS algorithm does prove useful for calculating RCSs, it may do so by using a finer mesh than classical algorithms do but solving the system to somewhat lower relative numerical precision, because that trade-off (higher mesh resolution but lower numerical precision) plays to the CJS algorithm's comparative advantages.

It is interesting to consider how the trade space of spatial resolution versus numerical precision plays out in practice for the overall accuracy of RCS calculations and just how much additional spatial resolution is required to make up for a decrease in numerical precision.[72] For example, lower-frequency radars (less than 1 GHz) are becoming increasingly common in military surveillance applications because of their longer ranges and advances that have been made in signal processing that overcome some of the radars' shortfalls.[73] At lower frequencies, the FEM mesh can be physically larger, which reduces $N$ for a given problem size. If, however, we are focused on higher frequencies, such as those increasingly used for missile seekers, automotive radars, or potentially for space-based collision avoidance sensors (those above, perhaps, 20 GHz),[74] then the wavelengths at play are much smaller, leading to increases in $N$. There is no one solution for all problem sets, and radio frequency engineers should be engaged to identify the situations in which trading numerical precision for spatial resolution may be advantageous.

## Sparsity of the *A* Matrix for Various Meshes

One main takeaway from the previous subsection is that the runtime of the CJS algorithm is extremely sensitive to the sparsity parameter $d$; we want this value to be as low as possible for the CJS algorithm to be useful. As we discussed previously, the sparsity of the *A* matrix is determined by the topology of the spatial mesh for our simulation. Each finite element contributes a nonzero matrix entry for only the mesh edges to which it is adjacent.

Figure 5.1 illustrates how this works with a simple example: the regular square mesh that Scherer et al. considered.[75] Consider, for example, the thick edge in the center of the figure labeled 4. (There is nothing special about this edge within the mesh; we have highlighted it only for visual clarity.) This edge is adjacent to two finite elements (i.e., squares), labeled A and B. Recall that the rows and columns of the *A* matrix correspond with the edges of this mesh. The finite element A contributes nonzero entries to the $A$ matrix between the edges numbered 1, 2, 3, and 4 ($4^2 = 16$ entries total). The finite element B contributes nonzero entries between the edges numbered 4, 5, 6, and 7 (another 16 entries, but note that both finite elements A and B contribute to the matrix entry $A_{4,4}$ because they both share edge 4, so these two finite elements contribute only 31 nonzero entries in total). If we consider the row of the matrix contributing to edge 4, we see that it has seven nonzero entries: $A_{4,1}, A_{4,2}, A_{4,3}, A_{4,4}, A_{4,5}, A_{4,6}$, and $A_{4,7}$. Similarly, every other row of the matrix also contains seven nonzero entries,[76] so the sparsity parameter for this mesh is $d = 7$.

---

[72] In economics terms, we would describe this as the marginal rate of substitution between spatial resolution and numerical precision.

[73] Jianqi Wu, *Advanced Metric Wave Radar*, Springer, 2015.

[74] Marta Martí-Marqués, *Space-Based Radar System for Geostationary Debris Detection and Tracking at MEO*, International Astronautical Federation, IAC-05-B6.1.03, 2005.

[75] Artur Scherer, Benoît Valiron, Siun-Chuon Mau, Scott Alexander, Eric van den Berg, and Thomas E. Chapuran, "Concrete Resource Analysis of the Quantum Linear-System Algorithm Used to Compute the Electromagnetic Scattering Cross Section of a 2D Target," *Quantum Information Processing*, Vol. 16, No. 60, 2017.

**Figure 5.1. Illustration of How the Mesh Topology Determines the Sparsity Parameter $d$**

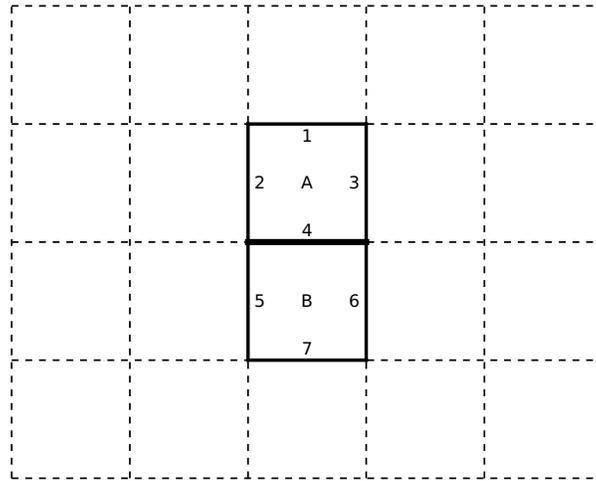

In general, we can calculate the sparsity parameter for the $A$ matrix corresponding with a given mesh topology using the following two-step process:

- For each edge $i$ of the mesh, count how many edges (including edge $i$ itself) are separated from edge $i$ by only a single volume element.
- Take the maximum of this count over all edges of the mesh.[77]

As we mentioned previously, the CJS algorithm must use a regular mesh to deliver an exponential speedup. Given how sensitive the runtime of the CJS algorithm is to the value of the sparsity parameter, the topology of this mesh is very important. For example, in the two-dimensional case, the triangular mesh illustrated in Figure 5.2 has a sparsity parameter $d = 5$ that is smaller than the $d = 7$ for the square mesh. It is somewhat more complicated to discretize the original differential equation onto the less symmetric triangular mesh, but this needs to be done only once, independently of the numerical calculation. And the payoff for the lower sparsity parameter is substantial: The $d^7$ dependence of the CJS algorithm's runtime on the sparsity parameter means that, holding all other parameters equal, going from $d = 7$ to $d = 5$ reduces the runtime by 91 percent.

---

[76] Except for the rows corresponding with the edges of the entire grid, which have only four nonzero entries.

[77] For a regular mesh, such as the one illustrated in Figure 5.1, the count will be the same for all edges of the mesh (except for the edges on the outer boundary of the mesh, for which the count will be smaller), so we can just pick any edge in the bulk of the mesh and skip step 2.

**Figure 5.2. Triangular Mesh Topology with Lower Sparsity Parameter Than the Square Topology**

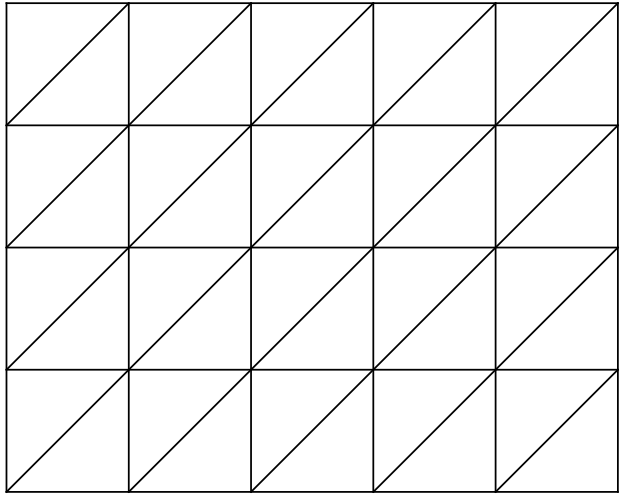

In the more-realistic case of three spatial dimensions, the mesh topology is even more important, because the sparsity parameters $d$ are much larger. Unfortunately, the strong sensitivity of the runtime on $d$ means that the CJS algorithm will run much more slowly on three-dimensional meshes than on two-dimensional meshes with a similar number of edges (with a much worse slowdown in going from two to three dimensions than the classical algorithms experience). For example, a bit of geometry shows that a cubic mesh has sparsity parameter $d = 33$.[78] A factor of $33^7 \approx 42$ billion in the runtime is very challenging to overcome, even with an exponential speedup in $N$.

For a fixed number of three-dimensional mesh edges, the $A$ matrix will be sparsest (and so the CJS algorithm will run fastest) on a tetrahedral mesh, because a tetrahedron is the volume element (i.e., polyhedron) with the lowest number of edges. Tetrahedral meshes are often used in finite-element analysis. Unfortunately, it is mathematically impossible to tile three-dimensional space with *regular* tetrahedra.[79] The best we can do is to regularly tile together irregular tetrahedra. For example, we can start with a cubic tiling and then subdivide each cube into six congruent (but irregular) tetrahedra. Alternatively, we could subdivide each cube into only five tetrahedra, but they cannot all be congruent. Given the runtime's strong sensitivity to the sparsity parameter, the latter option is probably better, even at the cost of an inelegant and less symmetric discretization mesh.

---

[78] Each edge in a cubic mesh is adjacent to four cubes arranged in a square, and those four cubes collectively have a total of 33 distinct edges when we take into account that some edges are shared between adjacent cubes.

[79] J. H. Conway and S. Torquato, "Packing, Tiling, and Covering with Tetrahedra," *Proceedings of the National Academy of Sciences*, Vol. 103, No. 28, July 11, 2006.

# Crossover Point Between the Conjugate Gradient and Clader-Jacobs-Sprouse Algorithms

Figure 5.3 illustrates how the runtimes of the classical CG and the quantum CJS algorithms (Equations 5.1 and 5.2) grow with problem size for the parameters $d = 7, \epsilon = 10^{-2}$ that Scherer et al. considered.[80] (Note that this is a linear-linear plot.) For simplicity, we assume that $C_{CG} = C_{CJS}$. This assumption is not very realistic—as we will discuss in the following subsection, $C_{CJS}$ is probably much bigger than $C_{CG}$ in practice—but it does not qualitatively affect the picture.

Figure 5.3. Growth in Runtime of the Conjugate Gradient (Blue) and Clader-Jacobs-Sprouse (Orange) Algorithms with System Size *N*

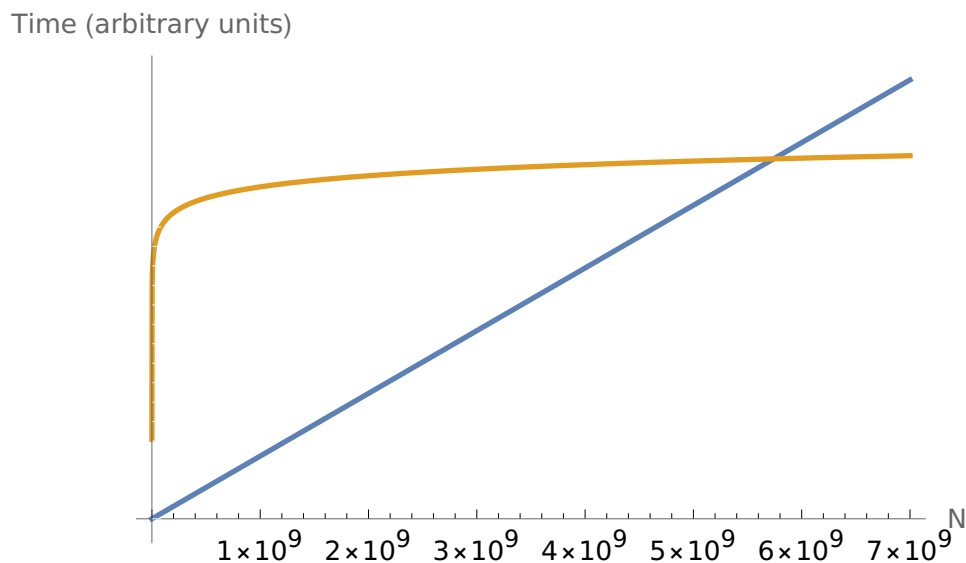

For these parameters, the terms involving $d$ and $\epsilon$ equal $8 \times 10^9$ for the CJS algorithm but only about 32 for the CG algorithm because of the latter algorithm's much better scaling in both parameters. So the logarithmic growth of the CJS algorithm has a relatively huge prefactor compared with the linear growth of the CG algorithm, and $N$ must become very large before the CJS algorithm becomes faster. The runtime of the CJS algorithm becomes very long very quickly for modest system sizes, then it levels out and grows much more slowly over a very large range of system sizes; the CG algorithm grows at a steady rate over all system sizes.

We can find the crossover value $N^*$ at which the CJS algorithm becomes faster than the CG algorithm by setting Equations 5.1 and 5.2 equal. The smaller the value of $N^*$, the more likely the CJS algorithm is to be practically useful. We find that $N^*$ satisfies the equation

$$\log N^* = -x\, N^*, \qquad (5.3)$$

where the parameter

---

[80] Scherer et al., 2017.

$$x = -\frac{C_{CG}}{C_{CJS}} \frac{\epsilon^2 \log(1/\epsilon)}{d^6}. \tag{5.4}$$

Note that $x$ is negative for $\epsilon \in (0, 1)$, and $x \to 0^-$ in the "classical-friendly" regime $\epsilon \ll 1$ or $d \gg 1$.

Equation 5.3 gives that

$$-\log N^* = xN^*,$$

$$\frac{1}{N^*} = e^{xN^*},$$

$$x = xN^* e^{xN^*}.$$

This is the defining equation for the Lambert W function, the special function defined as the inverse of the function $f(w) = we^w$. The solution of the last equation above is $xN^* = W(x)$ or $N^* = W(x)/x$.

The Lambert W function $W(x)$ is only real for $x \geq -1/e = -0.368...$, where $e$ (not to be confused with the error parameter $\epsilon$) denotes Euler's constant. If $x > -1/e$, then the asymptotic runtimes for the CJS and the CG algorithms intersect at two different values of $N$, as in Figure 5.3, and have one intersection at $N < e$ (not easily visible in figure). If $x = -1/e$, then the two curves become tangent at $N = e$. If $x < -1/e$, then the asymptotic quantum CJS algorithm runtime is less than the asymptotic classical CG runtime for all values of $N$.[81]

The Lambert W function $W(x)$ has two branches for $x \in [-1/e, 0)$. We are interested in the larger crossing point $N^*$, where the classical CG algorithm becomes permanently slower than the quantum CJS algorithm, so we want the branch that is more negative in value. This is the non-principal branch $W_{-1}(x)$. So we get the final formula

$$N^* = \frac{W_{-1}(x)}{x} \tag{5.5}$$

for the crossover value of $N$, with the parameter $x$ defined in Equation 5.4.

In the "classical-friendly" regime $d \gg 1$ or $0 < \epsilon \ll 1$, we have that $-1 \ll x < 0$. In this limit, the Lambert W function has the asymptotic form $W_{-1}(x) \sim \log(-x)$ as $x \to 0^-$, so

$$N^* \sim \frac{\log(-x)}{x} \text{ as } x \to 0^-. \tag{5.6}$$

Therefore, neglecting subleading logarithmic corrections, in the "classical-friendly" regime where $-1 \ll x < 0$, the crossover value $N^*$ is approximately

---

[81] This does not mean that the CJS algorithm actually runs faster than the CG algorithm for all $N$. The asymptotic Equations 5.1 and 5.2 apply only for large $N$, so we cannot use them to draw conclusions about the runtimes at small $N$.

$$N^* \approx -\frac{1}{x} = \frac{C_{CJS}}{C_{CG}} \frac{d^6}{\epsilon^2 \log(1/\epsilon)}. \tag{5.7}$$

We can use the approximate Equations 5.6 and 5.7 to make rough order-of-magnitude estimates for the crossover value without needing to use the obscure Lambert W function.

We check these approximations for the parameters that Scherer et al. considered in their paper:[82] $d = 7$ and $\epsilon = 0.01$. In this case, $x = -3.91 \times 10^{-9} \times \frac{C_{CG}}{C_{CJS}}$. We would expect that the slower qubit gate speeds and the overhead required for quantum error correction would generally make $C_{CJS}$ significantly larger than $C_{CG}$ when measured in "wall-clock time," but let us be very optimistic and assume that $C_{CG} \approx C_{CJS}$. Even under this optimistic assumption, $|x| \sim 10^{-9} \ll 1$ is very deep into the "classical-friendly" regime, where we would expect the crossover value $N^*$ to be very large. In this case, Equation 5.5 gives that the exact crossover value $N^* = 5.74 \times 10^9$. The approximation given by Equation 5.6 yields the quite similar value $N^* \approx 4.95 \times 10^9$. The rougher approximation in Equation 5.7 gives the significantly less accurate value $N^* \approx 2.66 \times 10^8$, which is off by an order of magnitude.

## Conclusion

If we drop the subleading logarithmic correction to the approximation in Equation 5.7, we get the *very* rough estimate that

$$N^* \approx \frac{C_{CJS}}{C_{CG}} \frac{d^6}{\epsilon^2}. \tag{5.8}$$

After all these approximations, this estimate is too rough to be quantitively accurate, but it gives the most important scaling of the crossover value: as $d^6/\epsilon^2$. The crossover is quite sensitive to the desired precision: Each order of magnitude of improved precision increases the crossover value $N^*$ by *two* orders of magnitude. And the crossover is *extremely* sensitive to the sparsity parameter $d$: Doubling the value of $d$ increases the crossover value $N^*$ by a factor of 64.

Therefore, as we discussed previously, the sparsity of the matrix A is extremely important for CJS-type algorithms; every reduction in the maximum number of elements per row makes the quantum CJS algorithm much more competitive against the classical CG algorithm. So it is very important to choose a mesh topology that makes $d$ as small as possible, such as the tetrahedral topology.

---

[82] Scherer et al., 2017.

Chapter 6

# Quantitative Resource Estimates for the Clader-Jacobs-Sprouse Algorithm

In this chapter, we turn to quantitative resource estimates for the CJS algorithm. We discuss the quantitative estimates from Scherer et al. and then extend them further.[83]

Quantitative resource estimates for quantum algorithms are highly technical and difficult to produce. Adding to the difficulty is the fact that quantum software and hardware are emerging technologies, and improvements to either may change resource estimates. The resource requirements also depend sensitively on assumptions about the hardware performance of the physical qubits, which is still unknown.[84]

Therefore, the quantitative estimates in this chapter should be taken with a large grain of salt. Improved implementations of the same underlying algorithm can drastically reduce resource estimates over time. For example, between 2012 and 2025, the estimated number of physical qubits required to factor 2048-bit RSA (Rivest-Shamir-Adleman algorithm) using Shor's algorithm dropped from over 1 billion qubits to under 1 million qubits.[85] These estimates did not assume any hardware improvements at all but improved only implementation of the same basic algorithm on the same hardware. The estimated resource requirements for the CJS algorithm may well see similar reductions over time.

## Summary of Scherer et al.'s Numerical Results

Scherer et al. chose the values $d = 7$ and $\epsilon = 10^{-2}$ for the concrete example that they worked out in detail. They began by roughly estimating the crossover point between the classical and quantum algorithms by equating the asymptotic runtimes, as we did in Chapter 5. As we showed in that chapter, the exact result for that estimate is $N^* = W_{-1}(x)/x$, and, for these choices of parameters, $x = -3.91 \times 10^{-9}$, so $N^* = 5.74 \times 10^9$. But Scherer et al. instead chose the significantly lower value $N = 332{,}020{,}680 = 3.32 \times 10^8$, which they claimed to be the crossover; we do not know why. In the two-dimensional toy model that they considered, this value of $N$ corresponds to discretizing the two-dimensional space into a $12{,}885 \times 12{,}885$ square grid with $N = 332{,}020{,}680 \approx 2 \times 12{,}885^2$ edges.

---

[83] Scherer et al., 2017. Throughout this chapter, we will refer to this publication as Scherer et al., and we will not include a reference each time.

[84] Edward Parker and Michael J. D. Vermeer, "Estimating the Energy Requirements to Operate a Cryptanalytically Relevant Quantum Computer," arXiv, arXiv:2304.14344, April 27, 2023.

[85] Craig Gidney, "How to Factor 2048 Bit RSA Integers with Less Than a Million Noisy Qubits," arXiv, arXiv:2505.15917, May 21, 2025.

Each edge represented a spatial length of 0.1 meters, so the entire modeled space represented 1.288 km × 1.288 km. The authors took the incident radar beam to have amplitude $E_0 = 1.0$ V/m and wavelength $\lambda = 1.0$ m and took the scatterer to be a metallic square with side length $L = 2\lambda = 2.0$ m at the center of the FEM grid.[86] They took the incident wave's propagation wave vector to be parallel to one side of the scatterer and its polarization to be parallel to the other side. They assumed that the (bistatic) receiver was directly opposite the wave source (relative to the scatterer) and had the same polarization direction as the incident wave. They estimated that the condition number $\kappa = O(N^{2/3})$ in three-dimensional and $O(N)$ in two-dimensional space, but they instead assumed the much smaller value $\kappa = 10^4$ provided by their research sponsor, the Intelligence Advanced Research Projects Activity, which, indeed, proved to be reasonably accurate numerically.[87]

For quantum computing specialists, we note that Scherer et al. considered only ideal logical qubit counts with all-to-all connectivity; they made no attempt to incorporate the (quite high) physical-qubit overhead required by quantum error correction.[88]

They performed two parallel versions of their estimates, one version excluding and the other version including the computational resources required by the oracles.

### Excluding the Oracles

The first version of Scherer et al.'s estimate computed the resources required without the oracles. That is, this version of the estimate made the following two simplifying assumptions:

1. The estimate assumed that the quantum computation begins with the data qubits in an initial quantum state $|b\rangle$ encoding the classical vector $\boldsymbol{b}$ in the linear system $A\boldsymbol{x} = \boldsymbol{b}$. Stated differently: It assumed that the quantum state $|b\rangle = \hat{O}_b|0\rangle$ is initialized from the all-zero state via some unknown initialization oracle $\hat{O}_b$ whose computational resource requirements are ignored. In the language of the logic flow of the QLSA, discussed after Figure 4.1, this first assumption corresponds to skipping step 1.
2. This version of Scherer et al.'s estimate also assumed that the (nonzero) elements of the matrix $A$ are returned by a black-box matrix oracle that inputs the integers $i$ and $j$ and returns the matrix element $A_{ij}$. The estimate ignored any computational resources required to perform this calculation. In the language of the logic flow of the QLSA, this second assumption corresponds to skipping step 2.

---

[86] The authors claim that this made the scatterer a "200 × 200 square area of vertices," (Scherer et al., 2017, p. 59) but these numbers are not internally consistent; if $\lambda = 1.0$ m, each edge represented a length of 0.1 m, and the scatterer's side length $L = 2\lambda$ as they claimed, then the scatterer would be 20 × 20 in vertices, not 200 × 200 as they claimed. We did not attempt to resolve this discrepancy because it was not important for this report.

[87] As we discussed in Chapter 5 in the context of asymptotic runtimes, in general, the CJS algorithm requires a preconditioner to bound the condition number to be independent of $N$. Scherer et al. (2017) did not actually implement the preconditioning step; instead, they simply assumed the value $\kappa = 10^4$ per their sponsor's guidance. Therefore, their resource estimates may be optimistic because the preconditioning step may be necessary in general.

[88] Technical note: For their set of elementary logic gates, Scherer et al. (2017) chose the six single-qubit logic gates Pauli-X, Pauli-Y, Pauli-Z, Hadamard (H), phase shift (S), and $\pi/8$ phase shift (T), as well as the two-qubit logic gate CNOT (usually called XOR in the classical setting).

So this version of Scherer et al.'s estimate considered only steps 3 and 4 in the QLSA logic flow. As we mentioned in Chapter 4, these simplifying assumptions are common in the QLSA literature.

Without the oracles, Scherer et al.'s estimated qubit count was quite encouraging. They estimated that calculating the RCS to the desired precision would require only 341 logical qubits. Although this is still far beyond the capabilities of our existing quantum computer hardware, this is quite modest as quantum algorithms go. By comparison, using Shor's algorithm to factor 2048-bit RSA requires 1,399 logical qubits (under certain reasonable assumptions)—over four times as many.[89]

Scherer et al.'s estimated runtime, $3.30 \times 10^{25}$ time steps, was less encouraging. Shor's algorithm requires only $6.9 \times 10^{13}$ time steps. Scherer et al. also estimated that running the CJS algorithm would require $1.29 \times 10^{25}$ Toffoli gates compared with only $6.5 \times 10^9$ Toffoli gates for Shor's algorithm on 2048-bit RSA.[90] Even without the oracles, the CJS algorithm would take around 12 orders of magnitude longer to run than Shor's algorithm. We will discuss why later in this chapter.

## Including the Oracles

Scherer et al. then estimated the computational resource costs to implement the initialization oracle and the matrix oracle.[91] The results were ominous: The required number of logical qubits jumped up by nearly six orders of magnitude to $3 \times 10^8$—far more than the number that Shor's algorithm requires. The runtime increased to $1.8 \times 10^{29}$ time steps, and the number of required Toffoli gates increased to $9.5 \times 10^{28}$.

Therefore, assuming that the oracles required negligible computational resources turned out *not* to be a safe assumption. Scherer et al.'s analysis implies that the oracles actually dominate the overall computational resources; steps 1 and 2 of the QLSA algorithm, which most previous literature had ignored, turn out to be the computational bottleneck in practice.

## Conclusion

The punchline of Scherer et al.'s analysis is that implementing the CJS algorithm at a useful problem size would take vastly more resources than Shor's algorithm and would probably be infeasible on any conceivable hardware implementation.

---

[89] Gidney, 2025.

[90] Gidney, 2025. A *Toffoli gate* is a primitive logic gate on three qubits that is the quantum equivalent to a *controlled-controlled-not* gate: It flips the value of the third input qubit if and only if the first two input qubits take value 1. It is the hardest primitive logic gate to implement on certain quantum computer architectures, so counting the number of times that it must be implemented is a common proxy for the total computational resource costs.

[91] Scherer et al. (2017) also estimated the costs of the third oracle mentioned in the overview of the CJS algorithm in Chapter 4.

## Extrapolating to Other Problem Sizes

Equation 5.2 gives the asymptotic form of the runtime $T_{CJS} \sim C_{CJS} \kappa\, d^7\, \epsilon^{-2} \log N$. We can plug the concrete numerical values $d = 7, \epsilon = 10^{-2}, N = 3.32 \times 10^8, \kappa = 10^4, T_{CJS} = 1.8 \times 10^{29}$ from Scherer et al. into this asymptotic formula to estimate the value of the constant prefactor $C_{CJS}$, which would allow us to extrapolate the runtime out to arbitrary problem sizes. Of course, fitting a model parameter to a single data point is always dangerous. Even if we assume that Scherer et al.'s numerical estimates are indeed optimal—which they may not be—the problem size $N = 3.32 \times 10^8$ that they considered may not yet be out in the asymptotic regime where we can neglect subleading corrections.

Moreover, as we discuss in the next section, Scherer et al. call into question whether that asymptotic form is exactly correct. If it is not, then that would completely invalidate our estimate.

Noting these important caveats, we get the following rough estimate,

$$C_{CJS} \approx 1.11 \times 10^{14} \text{ logical time steps,} \tag{6.1}$$

which is enormously large. We find the discouraging result that just the *numerical prefactor* in the asymptotic runtime for the CJS algorithm, excluding all the problem-instance-specific parameters, is already larger than the *entire* runtime to factor RSA-2048 via Shor's algorithm.

## Wall-Clock Time Estimates

We can attempt to make this result more concrete by converting abstract logical time steps to actual *wall-clock* time (e.g., measured in seconds, days, years). To do so, we must leave the land of purely algorithmic analysis and make some assumptions about the notional quantum computer's hardware speed. Specifically, we must estimate its *logical clock rate*, which captures the minimum time required for the processor to perform one elementary logic operation.

This is yet another challenging estimate to make. We have very little idea of the hardware architecture of an eventual fault-tolerant quantum computer capable of executing an algorithm of this complexity. We do not even know the basic physical form of its elementary qubits, known as the *qubit modality*. Many qubit modalities are being developed, and their logic gate operation times span several orders of magnitude. One of the fastest qubit modalities, superconducting qubits, can implement two-qubit logic gates in about 45 nanoseconds,[92] while one of the slowest modalities, trapped ions, requires 600 microseconds—more than 10,000 times as long.[93]

But any fault-tolerant quantum computer capable of running the CJS algorithm will require extensive quantum error correction, which will make the *logical* clock rate much slower than the physical gate operation time. Because only small steps toward quantum error correction have been demonstrated experimentally, it is difficult to estimate this logical clock rate. Moreover, Scherer et al. did not incorporate any quantum error correction requirements in their resource estimate.

---

[92] Dongxin Gao, Daojin Fan, Chen Zha, Jiahao Bei, Guoqing Cai, Jianbin Cai, Sirui Cao, Xiangdong Zeng, Fusheng Chen, Jiang Chen, et al., "Establishing a New Benchmark in Quantum Computational Advantage with 105-Qubit Zuchongzhi 3.0 Processor," arXiv, arXiv:2412.11924, December 16, 2024.

[93] IonQ, "IonQ Aria: Practical Performance," webpage, last updated January 8, 2025.

As a *very* crude estimate, we will assume that the logical clock rate may be similar to one particular quantum computer implementation that has been studied in great detail in the previous literature: Gidney's analysis of a hypothetical quantum computer that implements the surface error–correction code on superconducting-transmon qubits to implement Shor's algorithm to factor a 2048-bit number.[94] Roughly speaking, that analysis concluded that when all the requirements of quantum error correction were included, one logical time step would last about 25 microseconds, implying an effective logical clock rate of 40 kHz. This is much slower than a conventional CPU, which is typically in the GHz range.[95]

Gidney's hypothetical quantum computer was very carefully tailored to specifically implement Shor's algorithm on 2048-bit numbers, so it could certainly not be easily adapted to implement the CJS algorithm.[96] Nevertheless, we will use that estimate for lack of any better one while reiterating that the estimate should be taken with a very large grain of salt. If each logical time step requires 25 microseconds, then the constant prefactor for wall-clock runtime becomes

$$C_{\text{wall-clock}} \approx (1.11 \times 10^{14} \text{ logical time steps}) \times \left(25 \frac{\text{microseconds}}{\text{time step}}\right) \approx 88 \text{ years.}$$

As a reminder, to convert this constant prefactor to an actual runtime, one needs to multiply it by $\kappa\, d^7\, \epsilon^{-2} \log N$, which equals $1.62 \times 10^{15}$ for Scherer's toy problem. So, the total algorithm runtime would be about 140 quadrillion years, or 10 million times the current age of the universe.

# Why Is the Clader-Jacobs-Sprouse Algorithm So Much Slower Than Other Quantum Algorithms That Deliver an Exponential Speedup?

The huge prefactor in Equation 6.1 may seem quite counterintuitive. The asymptotic runtimes of most practical computer algorithms do not have such enormous constant prefactors. And the CJS algorithm offers an asymptotic exponential speedup over the fastest known classical algorithm, just like Shor's algorithm and Lloyd's quantum simulation algorithm do. But the latter algorithms have predicted runtimes that can be long but are often within a human lifetime. Why does the CJS algorithm seem to be so much slower than other quantum algorithms that give a similar asymptotic speedup?

---

[94] Gidney, 2025.

[95] The most-complicated arithmetic operation that Gidney's (2025) hypothetical quantum computer would perform is adding 33-bit numbers, which would take it about 2 milliseconds. By comparison, the 1945 ENIAC computer could read, write, or add ten-digit numbers in about 200 microseconds. By a convenient coincidence, a ten-digit number has almost exactly the same maximum size as a 33-bit number, so it is reasonably accurate to say that that Gidney's hypothetical quantum computer would run only one-tenth as fast as ENIAC did.

[96] Technical note: We are sweeping many important nuances under the rug in making this rough comparison. For example, Gidney assumed a surface code cycle time of 1 microsecond and a surface code distance of 25, which multiply out to 25 microseconds per lattice surgery. The code distance of 25 was reverse-engineered to deliver a logical error rate of $10^{-15}$, which is sufficient for Shor's algorithm on a 2048-bit number. The CJS algorithm requires a much deeper logical circuit depth than Shor's algorithm, so it may well require a much larger code distance and therefore a longer logical clock cycle than Gidney's computer would. The required code distance scales only logarithmically in the desired logical error rate, but the additional error correction overhead would probably still be very significant in practice. But estimating it is far beyond the scope of this report.

As we discussed previously, Scherer et al.'s resource estimates for logical qubits and circuit depth requirements for the CJS algorithm—which are directly related to quantum hardware and runtime requirements, respectively—depend heavily on whether the resource burden of oracles is included in the resource estimate. These oracles account for most of the resource requirements estimated by Scherer et al., and *Hamiltonian simulation* subroutines, defined in the next paragraph, stand out as the primary source of oracle resource use. But modifications to the CJS algorithm have the potential to alleviate bottlenecks, such as the number of oracle calls and lack of parallelism, so the oracle resource burdens may improve with time.

A *Hamiltonian* is a theoretical object from physics that describes the time evolution of a quantum mechanical system. If $\hat{H}$ is a Hamiltonian, then the famous Schrödinger equation associates the time evolution of a quantum mechanical system with the exponential of the Hamiltonian, denoted by $e^{-it\hat{H}}$.[97] The variable $t$ represents time in the quantum mechanical system, and $i$ is the imaginary unit. Several of the subroutines of the CJS algorithm involve representing the physical RCS problem in terms of $e^{-it\hat{H}}$ for an appropriately chosen Hamiltonian. This representation is known as *Hamiltonian simulation*.[98]

There is a vast literature about implementing Hamiltonian simulation; this literature largely uses a fundamental process known as *Trotterization*,[99] which is the observation that if $A$ and $B$ are Hamiltonians, then the following identity approximately holds for large numbers $r$:

$$e^{A+B} \approx \left(e^{\frac{A}{r}}e^{\frac{B}{r}}\right)^r.$$

The size of $r$ is a key constraint to the resource efficiency of the CJS algorithm. Scherer et al. estimate that more than 99 percent of the logical qubit requirements for executing the CJS algorithm are committed toward Hamiltonian simulation and Trotterization because of a large $r$ requirement.

Results from Berry et al. show that the necessary size of $r$ is not sensitive to the problem size $N$—the dimension of the discretized system of Maxwell equations, in the case of RCS—but is positively correlated with the condition number $\kappa$.[100] Structural requirements of the CJS algorithm also force significantly large numbers of repetitions of applications of Hamiltonian simulation without any obvious possibility of parallel computation, effectively multiplying the resource burden from simulation.

---

[97] Technical note: We note that $\hat{H}$ is a self-adjoint operator on the Hilbert space of quantum states and $e^{-it\hat{H}}$ is a unitary operator on that Hilbert space. We have dropped the unimportant factor of Planck's constant $\hbar$.

[98] Technical note: The rough intuition for why the CJS algorithm requires Hamiltonian simulation is as follows. A linear system of equations $Ax = b$ can easily be adapted to a closely related system in which the matrix $A$ is self-adjoint. But a key subroutine of many powerful quantum algorithms, the quantum Fourier transform, works best with matrices that are unitary because, in that case, we can use the quantum Fourier transform to apply the quantum phase estimation algorithm. So to be able to apply the quantum Fourier transform to a self-adjoint matrix $\hat{A}$, we need to first convert it to a unitary matrix that encodes the same information. The matrix exponential $e^{-i\hat{A}t}$ (where $t$ is a continuous real parameter) achieves this goal. The Hamiltonian simulation subroutine is essentially a mathematical technique (drawing inspiration from physics) for efficiently mapping a sparse self-adjoint matrix $\hat{A}$ to its matrix exponential $e^{-i\hat{A}t}$.

[99] Masuo Suzuki, "Fractal Decomposition of Exponential Operators with Applications to Many-Body Theories and Monte Carlo Simulations," *Physics Letters A*, Vol. 146, No. 6, June 4, 1990.

[100] Dominic W. Berry, Graeme Ahokas, Richard Cleve, and Barry C. Sanders, "Efficient Quantum Algorithms for Simulating Sparse Hamiltonians," *Communications in Mathematical Physics*, Vol. 270, No. 2, March 2007.

Assuming that target precision and other characteristics of the physical RCS problem remain constant, these estimates imply that improvements to the CJS algorithm's query complexity for Hamiltonian simulation or fundamental improvements to Hamiltonian simulation methods would reduce the logical qubit count. Because Scherer et al. estimated that only hundreds of logical qubits would be required to run the CJS algorithm without oracles, dramatic improvements in either direction could bring the logical qubit requirement low enough to be feasible for quantum computers with robust error correction.

Indeed, Scherer et al. pointed out that in between the time when they performed their primary analysis and when they wrote their paper, researchers had already discovered several new and faster algorithms for Hamiltonian simulation;[101] Scherer et al. estimated that these new algorithms could reduce their estimated runtimes by five orders of magnitude.[102] Since Scherer et al. published their paper, researchers have made many further improvements in Hamiltonian simulation algorithm that incorporate a high degree of parallelism.[103] It is not clear whether the CJS algorithm can take advantage of these new, highly parallelized algorithms; this is an important area for future research. If the algorithm can take advantage, then that could knock many more orders of magnitude off of Scherer et al.'s estimated runtimes.

## More Details on Hamiltonian Simulation as the Bottleneck

A quantitative analysis of Scherer et al.'s results confirms that Hamiltonian simulation is the clear computational bottleneck. There are two reasons for this. First, the authors call their Hamiltonian simulation subroutine nearly 200,000 times sequentially. Roughly speaking, this number of subroutine calls scales as approximately $12\,\epsilon^{-2}$.[104] Second, each Hamiltonian simulation subroutine call discretizes the time parameter at an extremely fine resolution of $2.5 \times 10^{12}$ time steps. (The biggest contribution in the formula for that time resolution is $\kappa^{5/4}/\epsilon^{3/2} = 10^8$.) Multiplying these two quantities together, the authors' computation would require performing Hamiltonian simulation over $\sim 10^{18}$ sequential time steps in total.[105] This would require more than $10^{20}$ matrix oracle queries, and the matrix oracle circuit requires over $10^8$ logical time steps to run.

---

[101] Dominic W. Berry, Andrew M. Childs, Richard Cleve, Robin Kothari, and Rolando D. Somma, "Exponential Improvement in Precision for Simulating Sparse Hamiltonians," *STOC '14: Proceedings of the Forty-Sixth Annual ACM Symposium on Theory of Computing*, May 31, 2014; Dominic W. Berry, Andrew M. Childs, and Robin Kothari, "Hamiltonian Simulation with Nearly Optimal Dependence on all Parameters," *2015 IEEE 56th Annual Symposium on Foundations of Computer Science*, 2015.

[102] Using our hardware estimates from the previous chapter, the algorithm would unfortunately still take 100 times the age of the universe to run, even with this improvement.

[103] Joonho Lee, Dominic W. Berry, Craig Gidney, William J. Huggins, Jarrod R. McClean, Nathan Weibe, and Ryan Babbush, "Even More Efficient Quantum Computations of Chemistry Through Tensor Hypercontraction," *PRX Quantum*, Vol. 2, No. 3, July 8, 2021.

[104] This expression simplifies the exact formula by neglecting discretization effects and approximating $2^{\lceil \log_2 M \rceil} \approx M$. The exact scaling formula is more complicated and makes the result somewhat larger than this approximation.

[105] Note that the number of time steps that the Hamiltonian simulation subroutine simulates over is different from the number of logical time steps required to execute the circuit.

The matrix oracle is called $10^{15}$ times more often than the state preparation oracle, so the latter oracle makes a comparatively negligible contribution to the runtime. Interestingly, though, the state preparation oracle requires an order of magnitude more *memory* (i.e., number of qubits) than the matrix oracle and is the memory bottleneck for the overall algorithm. Because one oracle represents the runtime bottleneck and the other one represents the memory bottleneck, these two bottlenecks can probably be optimized fairly independently.

Scherer et al. attempted to explain the huge discrepancy between the asymptotic runtime estimates suggested by Equation 5.2 and their concrete resource estimates. They gave a detailed discussion, but, in our opinion, their most important claim was that Clader, Jacobs, and Sprouse's calculation of Equation 5.2 was incorrect and that the correct asymptotic expression has a stronger dependence on the parameters $d$, $\kappa$, and $\epsilon$.[106]

Scherer et al. performed a very simple sensitivity analysis by rerunning their analysis for $N = 24$ with all other parameters the same as before. They found a negligible difference in the runtime, confirming that its dependence on $N$ is indeed probably only logarithmic, as Equation 5.2 would suggest. They did not attempt to change any other parameters, so they did not test how the runtime depends on those other parameters. A natural topic for future research would be to perform a more careful asymptotic analysis of how the CJS algorithm depends on the other parameters.

Scherer et al. judged that the most important bottleneck in practice was the accuracy parameter $\epsilon = 10^{-2}$, which required the use of large registers. Notably, the CKS algorithm discussed in Chapter 4 provides an exponential improvement over the HHL subroutine in the runtime's dependence on $\epsilon$, suggesting that the CKS algorithm may be able to remove that bottleneck. Unfortunately, even if the CKS removes the Hamiltonian simulation bottleneck, it would not necessarily deliver an exponential speedup for the full CJS algorithm; it might simply shift the bottleneck over to another step in the algorithm. We have not investigated the situation in detail, but we believe that simply replacing the HHL subroutine in the CJS algorithm with the CKS algorithm (if doing so is even possible) would probably just improve the asymptotic runtime scaling Equation 5.2 to $\kappa\, d^7 \log(N) \log(1/\epsilon)\, /\epsilon$—a significant improvement over the $\epsilon^{-2}$ dependence of the original CJS algorithm but short of an exponential speedup.[107] Attempting to find an algorithm that gives an exponential speedup with respect to the accuracy $\epsilon$ is another promising direction for future research.

---

[106] More precisely, they argued that the $\tilde{O}()$ notation mentioned in Chapter 5 conceals some qualitatively important multiplicative corrections to the runtime.

[107] Technical note: Roughly speaking, at its core, the CJS algorithm consists of a Hamiltonian simulation subroutine nested inside an amplitude estimation subroutine (which itself combines the Grover search and quantum Fourier transform subroutines). See Figure 1 in the supplemental material for Clader, Jacobs, and Sprouse (2013) for a schematic illustration of how these various subroutines fit together.

The factor of $\epsilon^{-2}$ in the original CJS runtime in Equation 5.2 combines one factor of $\epsilon^{-1}$ from the inner Hamiltonian simulation subroutine and another factor of $\epsilon^{-1}$ from the outer amplitude estimation subroutine. Because these subroutines are nested, the two factors of $\epsilon^{-1}$ from their respective runtimes multiply together to yield $\epsilon^{-2}$. The CKS algorithm only speeds up the inner Hamiltonian simulation subroutine (exponentially). So implementing CKS would improve the inner factor of $\epsilon^{-1}$ to $\log(1/\epsilon)$ but would leave the outer factor of $\epsilon^{-1}$ unchanged, and the total runtime would scale as $\log(1/\epsilon)/\epsilon$.

But it is possible that the outer factor of $1/\epsilon$ could also be exponentially improved by using a tomography protocol more efficient than the amplitude estimation algorithm. It is sometimes possible to estimate certain observables from a quantum state using protocols that are more efficient than the amplitude estimation algorithm, such as the *shadow tomography* protocol. Investigating

### Note on Oracle Implementation

We also mention one technical point about Scherer et al.'s implementation. Their research sponsors provided them with the matrix oracle in the form of a quite complicated Matlab function that used many trigonometric functions, including the arctan function. Implementing these functions from scratch using only low-level logic is very challenging. The researchers had to convert these Matlab functions into the Haskell programming language by hand using fixed-point arithmetic, then they used the Quipper programming language to convert these Haskell templates into quantum logic circuits. The resulting oracle circuits were extremely large, resulting in high overhead from using oracles.

Scherer et al. stated their belief that hand-coding the oracles at a lower level could reduce their circuit size only by a factor of five to ten. But we are more optimistic on this point than they are. In our experience, many implementations of quantum algorithms can be made much more efficient through hand-optimization and the use of problem-specific shortcuts and controlled approximations.[108] We suspect that these oracle circuits could be made much smaller with a more careful problem-specific manual analysis.

## Conclusion

The key bottleneck of the CJS algorithm, Hamiltonian simulation, turns out to be mathematically identical to the problem of directly simulating the time evolution of a quantum system. This Hamiltonian simulation problem is one of the main proposed applications of quantum computers. In fact, it has been much more deeply studied than the CJS algorithm itself has. For example, many pharmaceutical companies are making significant investments in quantum computing research and development because of its potential to deliver high-value applications in drug discovery.[109] So our surprising conclusion is that the practical feasibility of the CJS algorithm may well depend on future progress in the distinct field of quantum simulation, which a priori may seem completely unrelated. As strange as it may seem, research conducted by pharmaceutical companies may well eventually lead to a breakthrough in the design of stealth aircraft. If this happened, it would nicely illustrate the unexpected interconnections among scientific fields and the difficulties of predicting the eventual applications of a line of scientific research.

---

whether efficient protocols, such as shadow tomography, can be applied to this problem is an important area of future research. Scott Aaronson, "Shadow Tomography of Quantum States," arXiv, arXiv:1711.01053, November 13, 2018; Hsin-Yuan Huang, Richard Kueng, and John Preskill, "Predicting Many Properties of a Quantum System from Very Few Measurements," *Nature Physics*, Vol. 16, No. 10, October 2020.

[108] Gidney, 2025.

[109] Anita Chandran, "Biopharma Foresees a 'Quantum Advantage': They Could Be Right," *Nature Biotechnology*, Vol. 42, No. 5, May 2024.

Chapter 7

# Discussion

The CJS algorithm is one of the very few quantum algorithms (along with Shor's factoring algorithm and Lloyd's algorithm for simulating quantum dynamics) that demonstrates an exponential speedup over the best known classical algorithm for a well-studied problem of practical importance. Therefore, on the basis of first principles, we believe that the CJS algorithm is one of the quantum algorithms with the most promise to become technically feasible to run on a quantum computer faster than the best corresponding classical algorithm can run on a classical computer.[110]

But, as we have discussed throughout this report, there is a fairly long list of caveats to this conclusion:

1. The CJS algorithm does not speed up all classical approaches to calculating RCSs. It speeds up only one particular approach: the FD-FEM. In RCS regimes in which other classical approaches are more useful than the FD-FEM method, the CJS algorithm does not necessarily give any advantage over classical approaches.[111]
2. One of the main advantages of the classical FEM is that it can be used on an arbitrary mesh. In particular, we can give the mesh finer resolution in areas where the fields are changing quickly, so higher precision is necessary. But the CJS algorithm can offer a speedup for the FEM only on a regular mesh, which removes one of the main advantages of the FEM.
3. The CJS algorithm gives an exponential speedup over the fastest known classical algorithm for calculating RCSs, but there is no proof that it gives an exponential speedup over the fastest possible classical algorithm. It is possible that someone could discover a classical algorithm that matches the performance of the CJS algorithm.[112] There is some precedent for this situation: Researchers have proposed quantum algorithms for other problems that were

---

[110] Saying that it will become *technically feasible* to run is not the same as saying that it will actually *be* run. The latter depends on economics.

Which quantum algorithms will actually be deployed first will depend on the demand side (i.e., the real or perceived societal benefit), as well as the supply side (i.e., the resource cost of deployment). Quantum simulation algorithms may have significant commercial applications, such as drug discovery, but the CJS algorithm will probably have mainly military applications around stealth. It is very difficult to compare the value of commercial research and development, which leads to measurable economic profits, with the value of military research and development, which does not. So we do not attempt to estimate which application will be more societally valuable.

But we note that slower quantum algorithms, such as Grover's algorithm, have a much broader range of applicability than the CJS algorithm does. So it is possible that Grover's algorithm will be deployed first because of higher demand despite its higher resource costs.

[111] Of course, the caveat to this caveat is that if the CJS algorithm hugely speeds up the FD-FEM method, then the quantum-accelerated FD-FEM method might become more broadly useful than it is today, and it could supplant classical approximate methods for certain problems.

[112] Aaronson, 2015.

temporarily believed to give an exponential advantage over the best classical algorithm, but then other researchers discovered new classical algorithms that matched the quantum algorithm's performance.[113] Nevertheless, the fact that the problem of calculating RCSs has been heavily studied for many decades but no one has publicly published any classical FD-FEM algorithms that improve on the CG method offers circumstantial evidence that a faster classical algorithm may not exist (or at least may be very difficult to find).

4. The CJS algorithm depends very strongly on the sparsity parameter $d$, as $d^7$. The sparsity parameter, in turn, depends on the topology of the spatial mesh, and three-dimensional meshes have quite high sparsity parameters. So for any three-dimensional mesh, the $d^7$ factor will make a very large contribution to the algorithm's runtime. The algorithm will therefore run much faster on a tetrahedral mesh than on any other mesh topology.
5. Concrete resource estimates from the literature suggest that the constant prefactor implicit in the asymptotic scaling expression is absolutely enormous. Combined with some reasonable hardware assumptions, these estimates imply that calculating an RCS for even a simple two-dimensional toy problem would require a runtime that is 10 million times the age of the universe. But these resource estimates did not incorporate the latest algorithmic developments, which could reduce the runtime by at least five orders of magnitude (and potentially many more).

These five major caveats lead us to conclude that the CJS algorithm is significantly less likely than either quantum simulation algorithms or Shor's algorithm to provide a practical quantum advantage in the foreseeable future.

Point 4 indicates that the topology of the spatial mesh is very important. In two-dimensional space, going from a square mesh to a triangular mesh with the same number of edges (with all other parameters held equal) speeds up the runtime by 91 percent. In three-dimensional space, the mesh topology is even more important. For the CJS to gain the maximum possible advantage on a real three-dimensional problem, the CJS algorithm should use a tetrahedral rather than a cubic mesh. Even though it is more complicated to discretize the Helmholtz differential equation on a tetrahedral mesh, the improvement in runtime will probably justify this one-time upfront cost.

We also found that if the CJS algorithm does become practical, it may trade off higher spatial resolution than classical algorithms in exchange for reduced numerical precision. This trade-off may be more advantageous at higher radar frequencies, where shorter wavelengths require higher spatial resolution.

The news is not all bad for the CJS algorithm. We have identified a very promising path for future research: exploring whether it is possible to replace the HHL subroutine within the CJS algorithm with a faster subroutine, such as the CKS algorithm. If the CJS oracles satisfy the technical requirements for the CKS algorithm, then CKS could qualitatively speed up the performance of the CJS algorithm. These improvements would not necessarily bring the algorithm's resource requirements down near those of more familiar applications, such as factoring or quantum chemistry, but further research is required to know for sure.[114]

---

[113] Samuel Greengard, "The Algorithm That Changed Quantum Machine Learning," *Communications of the ACM*, Vol. 62, No. 8, August 2019.

We found that the critical bottleneck for the CJS algorithm is the Hamiltonian simulation subroutine, which is much more often applied to the task of simulating actual physical quantum systems. Therefore, we reach the surprising conclusion that a breakthrough in the modeling of RCSs could well emerge from of a research effort with a completely different goal, such as medical drug discovery.

We have identified three important topics for future research:

1. As we mentioned previously, investigate whether recent improvements to the HHL algorithm (such as the CKS algorithm or even more-recent improvements in Hamiltonian simulation) might greatly speed up the CJS algorithm from its original version.
2. More carefully work out the asymptotic runtime that Clader, Jacobs, and Sprouse roughly calculated and better understand how it depends on the parameters $d$, $\kappa$, and $\epsilon$ other than the problem size $N$.[115]
3. More efficiently implement the CJS oracles (or approximate versions) as quantum circuits without going through the cumbersome pipeline of converting them from Matlab code to Haskell code to quantum logic circuits, which probably added unnecessary overhead.

It is inherently extremely difficult to estimate the resource requirements for a hypothetical computer that vastly exceeds the capabilities of any existing computer and whose basic architecture is still unknown. Therefore, all of our results come with enormous (but unquantifiable) error bars. Broadly speaking, throughout Chapters 5 and 6, we made a long chain of assumptions that yielded results that became increasingly concrete but probably less accurate. We encourage the reader not to take our estimate of 140 quadrillion years as the final conclusion of this report. We believe there are far too many uncertainties leading up to that estimate for it to be quantitatively accurate. Instead, we believe that each step in our chain of reasoning yielded interesting and useful intermediate conclusions.

---

[114] Technical note: As we discussed near the end of Chapter 6 in the context of Hamiltonian simulation as a computational bottleneck, the basic structure of the CJS algorithm—a Hamiltonian simulation subroutine nested inside an amplitude estimate routine—is (unfortunately) inherently difficult to speed up. The inner Hamiltonian simulation loop and the outer amplitude estimation loop both independently contribute a rather large factor of $\epsilon^{-1}$ to the runtime, so *both* loops need to be independently sped up exponentially to bring the total algorithm runtime down to polylogarithmic in $1/\epsilon$, at which point CJS would probably be competitive with other algorithms. Finding an exponentially speedup for the inner (Hamiltonian simulation) loop seems quite feasible, but speeding up the outer (amplitude estimation) loop may be more challenging.

[115] Technical note: The specific goal would be to replace the $\tilde{O}()$ asymptotic dependence implied by the tilde in Equation 5.2 to a tighter $\Theta()$ dependence that captures all the subleading multiplicative terms.

Appendix

# Discussion of Numerical Precision

In this appendix, we translate numerical precision from the units typically used in quantum computing literature (a multiplicative error term in linear units) to those typically used in the RCS engineering literature (a linear offset in decibels), and we relate those precision levels to common tasks for which RCS may be used.

The *total* error $\varepsilon$ in the exact RCS is a combination of two different errors: (1) the error from the spatial discretization of Maxwell's equations (controlled by the parameter $N$, the density of the mesh) and (2) the error in solving the resulting discretized linear system of equations (given by the parameter $\epsilon$ in Chapters 5 and 6; note the different typeface between $\varepsilon$ and $\epsilon$). The discretization error contribution is difficult to estimate quantitatively, but the error $\epsilon$ from solving the linear system gives a lower bound on the total error $\varepsilon$. In this appendix, we will only consider the total error $\varepsilon$.

To translate the error term $\epsilon$ into units familiar to RF engineers, we note that it refers to a multiplicative error of the form:

$$\sigma_{lin}(1 - \varepsilon) \leq \tilde{\sigma}_{lin} \leq \sigma_{lin}(1 + \varepsilon),$$

where $\sigma_{lin}$ is the true target RCS (at a given aspect angle and frequency) in linear units ($m^2$), and $\tilde{\sigma}_{lin}$ is the estimate returned from the CJS algorithm. We define $\Delta$ as the additive RCS error in dB units:

$$\Delta = |10 \log_{10} \sigma_{lin} - 10 \log_{10} \tilde{\sigma}_{lin}|.$$

Using this relationship, we can relate the multiplicative error term ($\varepsilon$) with the additive RCS error term ($\Delta$):

$$\Delta = 10 \log_{10}(1 + \varepsilon).$$

This is plotted in Figure A.1. Note that the curve is approximately linear on a log-log scale; an order-of-magnitude increase in the multiplicative error term $\varepsilon$ corresponds roughly to an order-of-magnitude increase in the additive dB error term $\Delta$.

**Figure A.1. Conversion of Multiplicative Error Term to Additive dB Error**

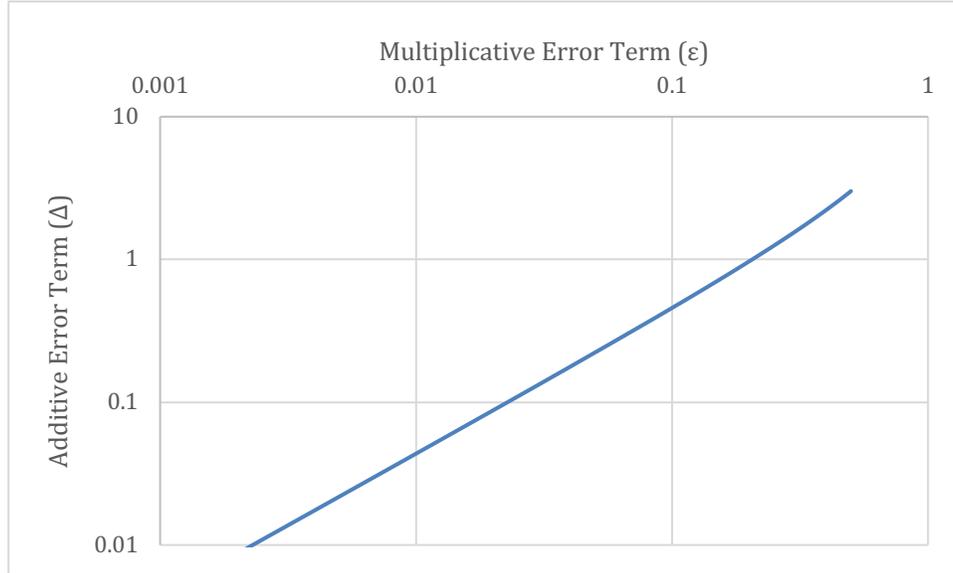

Table A.1 contains several data points from this relationship, with commentary on possible uses for the resulting RCS data and a calculation of the impact of RCS errors on radar detection range calculations (ignoring the impact of atmospheric loss).

**Table A.1. Selected RCS Error Regimes**

| Multiplicative Error ($\varepsilon$) | Additive Error ($\Delta$) | % Error in Radar Detection Range | Possible Uses |
|---|---|---|---|
| $10^{-4}$ | 0.00043 dB | <0.01% | |
| $10^{-3}$ | 0.0043 dB | 0.03% | Verification and validation[a] |
| $10^{-2}$ | 0.044 dB | 0.25% | Route planning[b] |
| $10^{-1}$ | 0.46 dB | 2.67% | Operational research[c] |

[a] *Verification and validation* refers to the certification that a system or capability is performing as required; in this case, we refer to predicted and delivered RCS for a flight article.
[b] *Route planning* refers to planning missions that depend on accurate understanding of the risk of detection and engagement by adversary radio frequency–based systems. A detection range error of less than 0.25 percent corresponds to an error of approximately 250 meters if the reported detection range of a system is 100 km. Errors larger than this will likely have a significant impact on operational planning.
[c] *Operational research* refers to supporting analysis for planning and acquisition, wherein notional scenarios are simulated to determine expected outcomes. Greater uncertainties in the future state of the world make highly accurate models impossible, and thus a reduced RCS precision is acceptable. Notionally, we claim in this case that RCS error of less than 0.46 dB, resulting in detection range errors on the order of 2.67 percent, is sufficient.

From this table, we conclude that setting the numerical precision to values less than 10 percent will be useful to radio frequency engineers in some form but that values less than 0.1 percent ($10^{-3}$) will be more broadly useful across a range of application areas.

To further illustrate the potential trade-off, we plot the asymptotic runtime estimates from Chapter 5 for the last three multiplicative error thresholds in Figure A.2. Unlike Figure 5.1, this plot is on a log-log scale. Route planning quality ($\varepsilon = 10^{-2}$) matches the curves shown in Chapter 5, while operations research quality ($\varepsilon = 10^{-1}$) shows a vastly reduced runtime (roughly two orders of magnitude) for large $N$ and reduces the crossover point from ~6 billion FEM edges to ~80 million FEM edges. However, if we wanted to replicate the precision needed for verification and validation of RCS measurements ($\varepsilon = 10^{-3}$), the startup cost of CJS would mean that it is not faster until the problem size exceeds ~500 billion FEM edges.

**Figure A.2. Impact of Numerical Precision on Crossover Point**

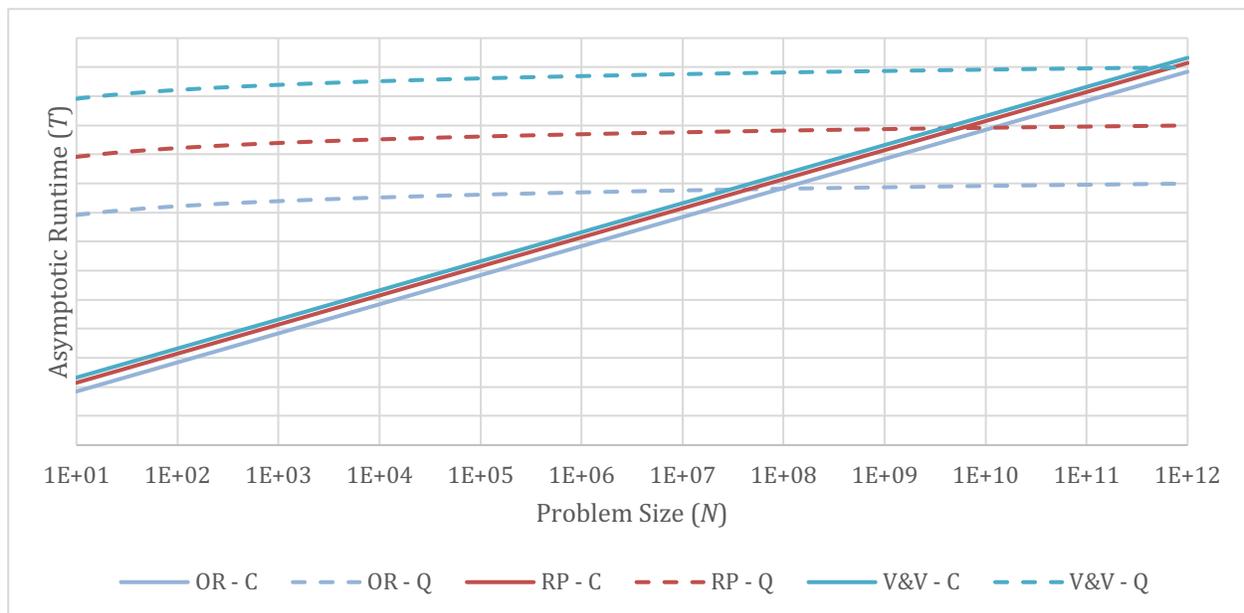

NOTE: OR = operational research quality ($\varepsilon = 10^{-1}$); RP = route planning quality ($\varepsilon = 10^{-2}$); V&V = verification and validation quality ($\varepsilon = 10^{-3}$); C = classical approach (CG); Q = quantum approach (CJS). Each horizontal or vertical gridline covers one order of magnitude.

These results, as with those shown in Chapter 5, assume a sparsity of $d = 7$, which is valid for two-dimensional problems on a square mesh. A three-dimensional mesh will have a larger value of $d$ (as high as 33 for a cubic mesh), leading to much larger problem sizes before the quantum algorithm overtakes the classical one.

# Abbreviations

| | |
|---|---|
| CG | conjugate gradient |
| CJS | Clader-Jacobs-Sprouse (algorithm) |
| CKS | Childs-Kothari-Somma (algorithm) |
| EM | electromagnetic |
| FD-FD | finite-difference frequency domain |
| FD-FEM | frequency-domain finite-element method |
| FD-TD | finite-difference time domain |
| FEM | finite-element method |
| HHL | Harrow-Hassidim-Lloyd (algorithm) |
| PDE | partial differential equation |
| QLSA | quantum linear system algorithm |
| RSA | Rivest-Shamir-Adleman (algorithm) |
| RCS | radar cross section |
| SDE | stochastic partial differential equation |

# References

Aaronson, Scott, "Read the Fine Print," *Nature Physics*, Vol. 11, No. 4, April 2, 2015.

Aaronson, Scott, "Shadow Tomography of Quantum States," arXiv, arXiv:1711.01053, version 2, November 13, 2018.

Alghassi, H., A. Deshmuk, N. Ibrahim, N. Robles, S. Woerner, and C. Zoufal, "A Variational Quantum Algorithm for the Feynman-Kac Formula," *Quantum*, Vol. 6, June 7, 2022.

Alhajjar, Elle, Jesse Geneson, Anupam Prakash, and Nicolas Robles, "Efficient Quantum Loading of Probability Distributions Through Feynman Propagators," arXiv, arXiv:2311.13702, version 2, November 28, 2023.

Ambainis, Andris, "Variable Time Amplitude Amplification and a Faster Quantum Algorithm for Solving Systems of Linear Equations," arXiv, arXiv:1010.4458, November 14, 2010.

Arute, Frank, Kunal Arya, Ryan Babbush, Dave Bacon, Joseph C. Bardin, Rami Barends, Rupak Biswas, Sergio Boixo, Fernando G. S. L. Brandao, David A. Buell, et al., "Quantum Supremacy Using a Programmable Superconducting Processor," *Nature*, Vol. 574, No. 7779, October 23, 2019.

Babbush, Ryan, Jarrod R. McClean, Michael Newman, Craig Gidney, Sergio Boixo, and Hartmut Neven, "Focus Beyond Quadratic Speedups for Error-Corrected Quantum Advantage," *PRX Quantum*, Vol. 2, No. 1, March 29, 2021.

Berry, Dominic W., Graeme Ahokas, Richard Cleve, and Barry C. Sanders, "Efficient Quantum Algorithms for Simulating Sparse Hamiltonians," *Communications in Mathematical Physics*, Vol. 270, No. 2, March 2007.

Berry, Dominic W., Andrew M. Childs, Richard Cleve, Robin Kothari, and Rolando D. Somma, "Exponential Improvement in Precision for Simulating Sparse Hamiltonians," *STOC '14: Proceedings of the Forty-Sixth Annual ACM Symposium on Theory of Computing*, May 31, 2014.

Berry, Dominic W., Andrew M. Childs, and Robin Kothari, "Hamiltonian Simulation with Nearly Optimal Dependence on All Parameters," *2015 IEEE 56th Annual Symposium on Foundations of Computer Science*, 2015.

Berry, Dominic W., Andrew M. Childs, Aaron Ostrander, and Guoming Wang, "Quantum Algorithm for Linear Differential Equations with Exponentially Improved Dependence or Precision," *Communications in Mathematical Physics*, Vol. 356, No. 3, December 2017.

Bramble, James H., Joseph E. Pasciak, and Jinchiao Xu, "Parallel Multilevel Conditioners," *Mathematics of Computation*, Vol. 55, No. 191, 1990.

Chandran, Anita, "Biopharma Foresees a 'Quantum Advantage': They Could Be Right," *Nature Biotechnology*, Vol. 42, No. 5, May 2024.

Chatterjee, A., J. M. Jin, and J. L. Volakis, "Edge-Based Finite Elements and Vector ABCs Applied to 3-D Scattering," *IEEE Transactions on Antennas and Propagation*, Vol. 41, No. 2, February 1993.

Childs, Andrew M., Robin Kothari, and Rolando D. Somma, "Quantum Algorithm for Linear Systems of Equations with Exponentially Improved Dependence on Precision," *SIAM Journal on Computing*, Vol. 46, No. 6, 2017.

Childs, Andrew M., Jin-Peng Liu, and Aaron Ostrander, "High-Precision Quantum Algorithms for Partial Differential Equations," *Quantum*, Vol. 5, November 10, 2021.

Cho, Edmond, "A Priori Sparsity Patterns for Parallel Sparse Approximate Inverse Preconditioners," *SIAM Journal on Scientific Computing*, Vol. 21, No. 5, 2000.

Clader, B. D., B. C. Jacobs, and C. R. Sprouse, "Preconditioned Quantum Linear System Algorithm," *Physical Review Letters*, Vol. 110, No. 25, June 2013.

Conway, J. H., and S. Torquato, "Packing, Tiling, and Covering with Tetrahedra," *Proceedings of the National Academy of Sciences*, Vol. 103, No. 28, July 11, 2006.

Deiml, Matthias, and Daniel Peterseim, "Quantum Realization of the Finite Element Method," *Mathematics of Computation*, August 14, 2025.

Feynman, Richard P., "Simulating Physics with Computers," *International Journal of Theoretical Physics*, Vol. 21, Nos. 6–7, June 1982.

Fontanela, Filipe, Antoine Jacquier, and Mugad Oumgari, "A Quantum Algorithm for Linear PDEs Arising in Finance," *SIAM Journal on Financial Mathematics*, Vol. 12, No. 4, 2021.

Gao, Dongxin, Daojin Fan, Chen Zha, Jiahao Bei, Guoqing Cai, Jianbin Cai, Sirui Cao, Xiangdong Zeng, Fusheng Chen, Jiang Chen, et al., "Establishing a New Benchmark in Quantum Computational Advantage with 105-qubit Zuchongzhi 3.0 Processor," arXiv, arXiv:2412.11924, December 16, 2024.

Gidney, Craig, "How to Factor 2048 Bit RSA Integers with Less Than a Million Noisy Qubits," arXiv, arXiv:2505.15917, May 21, 2025.

Gonzalez-Conde, Javier, Ángel Rodríguez-Rozas, Enrique Solano, and Mikel Sanz, "Efficient Hamiltonian Simulation for Solving Option Price Dynamics," *Physical Review Research*, Vol. 5, No. 4, December 8, 2023.

Greengard, Samuel, "The Algorithm That Changed Quantum Machine Learning," *Communications of the ACM*, Vol. 62, No. 8, August 2019.

Grote, Marcus J., and Thomas Huckle, "Parallel Preconditioning with Sparse Approximate Inverses," *SIAM Journal on Scientific Computing*, Vol. 18, No. 3, 1997.

Harrow, Aram W., Avinatan Hassidim, and Seth Lloyd, "Quantum Algorithm for Linear Systems of Equations," *Physical Review Letters*, Vol. 103, No. 15, October 9, 2009.

Henry, Justin K. A., Ram M. Narayanan, and Puneet Singla, "Radar Cross-Section Modeling of Space Debris," in Erik Blasch, Frederica Darema, and Alex Aved, eds., *Dynamic Data Driven Applications Systems: 4th International Conference, DDDAS 2022, Cambridge, MA, USA, October 6–10, 2022, Proceedings*, Springer Cham, 2024.

Hoefler, Torsten, Thomas Häner, and Matthias Troyer, "Disentangling Hype from Practicality: On Realistically Achieving Quantum Advantage," *Communications of the ACM*, Vol. 66, No. 5, May 2023.

Hu, Junpeng, Shi Jin, and Lei Zhang, "Quantum Algorithms for Multiscale Partial Differential Equations," arXiv, arXiv:2304.06902, April 14, 2023.

Huang, Hsin-Yuan, Richard Kueng, and John Preskill, "Predicting Many Properties of a Quantum System from Very Few Measurements," *Nature Physics*, Vol. 16, No. 10, October 2020.

IonQ, "IonQ Aria: Practical Performance," webpage, last updated January 8, 2025. As of September 5, 2025: https://ionq.com/resources/ionq-aria-practical-performance

Jackson, John David, *Classical Electrodynamics*, 3rd ed., Wiley, 1998.

Jin, Jian-Ming, *The Finite Element Method in Electromagnetics*, 2nd ed., Wiley, 2002.

Jin, Shi, Nana Liu, and Yu Yue, "Quantum Simulation of Partial Differential Equations via Schrödingerization," *Physical Review Letters*, Vol. 133, No. 23, December 6, 2024.

Jordan, Stephen P., "Quantum Algorithm Zoo," last updated March 31, 2025. As of March 31, 2025: https://quantumalgorithmzoo.org/

Kubo, Kenji, Yuya O. Nakagawa, Suguru Endo, and Shota Nagayama, "Variational Quantum Simulations of Stochastic Differential Equations," *Physical Review A*, Vol. 103, No. 5, May 21, 2021.

Ladd, T. D., F. Jelezko, R. Laflamme, Y. Nakamura, C. Monroe, and J. L. O'Brien, "Quantum Computers," *Nature*, Vol. 464, No. 7285, March 4, 2010.

Lee, Joonho, Dominic W. Berry, Craig Gidney, William J. Huggins, Jarrod R. McClean, Nathan Weibe, and Ryan Babbush, "Even More Efficient Quantum Computations of Chemistry Through Tensor Hypercontraction," *PRX Quantum*, Vol. 2, No. 3, July 8, 2021.

Liu, Yilian, *A Variational Quantum Algorithm for Solving Partial Differential Equations*, thesis, Cornell University, May 2023.

Lloyd, Seth, "Universal Quantum Simulators," *Science*, Vol. 273, No. 5278, August 23, 1996.

Martí-Marqués, Marta, *Space-Based Radar System for Geostationary Debris Detection and Tracking at MEO*, International Astronautical Federation, IAC-05-B6.1.03, 2005. As of August 18, 2025: https://iafastro.directory/iac/archive/browse/IAC-05/B6/1/1965/

Montanaro, Ashley, and Sam Pallister, "Quantum Algorithms and the Finite Element Method," *Physical Review A*, Vol. 93, No. 3, March 2016.

Morvan, A., B. Villalonga, X. Mi, S. Mandrà, A. Bengtsson, P. V. Klimov, Z. Chen, S. Hong, C. Erickson, I. K. Drozdov, et al., "Phase Transitions in Random Circuit Sampling," *Nature*, Vol. 634, No. 8033, October 9, 2024.

Parker, Edward and Michael J. D. Vermeer, "Estimating the Energy Requirements to Operate a Cryptanalytically Relevant Quantum Computer," arXiv, arXiv:2304.14344, April 27, 2023.

Prince, Todd, "Pentagon Faces 'Wake-Up Call' to Meet Drone Innovation Highlighted in Ukraine War," Radio Free Europe, August 7, 2025. As of August 18, 2025: https://www.rferl.org/a/pentagon-wake-up-call-drone-innovation-ukraine-war/33496970.html

Saad, Yousef, *Iterative Methods for Sparse Linear Systems*, Society for Industrial and Applied Mathematics, 2003.

Scherer, Artur, Benoît Valiron, Siun-Chuon Mau, Scott Alexander, Eric van den Berg, and Thomas E. Chapuran, "Concrete Resource Analysis of the Quantum Linear-System Algorithm Used to Compute the Electromagnetic Scattering Cross Section of a 2D Target," *Quantum Information Processing*, Vol. 16, No. 60, 2017.

Shor, Peter W., "Polynomial-Time Algorithms for Prime Factorization and Discrete Logarithms on a Quantum Computer," *SIAM Journal on Computing*, Vol. 26, No. 5, 1997.

Skolnik, Merrill I., ed., *Radar Handbook*, 3rd ed., McGraw Hill, 2009.

Suzuki, Masuo, "Fractal Decomposition of Exponential Operators with Applications to Many-Body Theories and Monte Carlo Simulations," *Physics Letters A*, Vol. 146, No. 6, June 4, 1990.

Wu, Jianqi, *Advanced Metric Wave Radar*, Springer, 2015.